\documentclass[pmlr]{jmlr}%

\usepackage{amsmath,amssymb,amsfonts}
\usepackage{graphicx}
\usepackage{lipsum}
\usepackage{float}
\usepackage{multirow}
\usepackage{ragged2e}
\usepackage{microtype}
\usepackage{enumitem}

\usepackage{longtable}%

 \usepackage{booktabs}
 \usepackage{adjustbox}
 \newcommand{\bftab}{\fontseries{b}\selectfont}

\makeatletter
\def\set@curr@file#1{\def\@curr@file{#1}} %
\makeatother

\theorembodyfont{\upshape}
\theoremheaderfont{\scshape}
\theorempostheader{:}
\theoremsep{\newline}

\jmlryear{2023}
\jmlrpages{}
\jmlrworkshop{Machine Learning for Healthcare}

\title[Neurological Prognostication Using Dynamic Survival Analysis]{Neurological Prognostication of Post-Cardiac-Arrest Coma Patients Using EEG Data: A Dynamic Survival Analysis Framework with Competing Risks}

\author{\Name{Xiaobin Shen} %
       \Email{xiaobins@andrew.cmu.edu}\\ 
       \addr Heinz College of Information Systems and Public Policy\\
       Carnegie Mellon University\\ \vspace{-0.8em}
       \\ 
       \Name{Jonathan Elmer} %
       \Email{elmerjp@upmc.edu}\\ 
       \addr Department of Emergency Medicine\\
       University of Pittsburgh\\ \vspace{-0.8em}
       \\
       \Name{George H. Chen} %
       \Email{georgechen@cmu.edu}\\ 
       \addr Heinz College of Information Systems and Public Policy\\
       Carnegie Mellon University\\
       }

\begin{document}

\maketitle

\begin{abstract}
Patients resuscitated from cardiac arrest who enter a coma are at high risk of death. Forecasting neurological outcomes of these patients (i.e., the task of \emph{neurological prognostication}) could help with treatment decisions: which patients are likely to awaken from their coma and should be kept on life-sustaining therapies, and which are so ill that they would unlikely benefit from treatment? In this paper, we propose, to the best of our knowledge, the first \emph{dynamic} framework for neurological prognostication of post-cardiac-arrest comatose patients using EEG data: our framework makes predictions for a patient over time as more EEG data become available, and different training patients' available EEG time series could vary in length. Predictions themselves are phrased in terms of either time-to-event outcomes (time-to-awakening or time-to-death) or as the patient's probability of awakening or of dying across multiple time horizons (e.g., within the next 24, 48, or 72 hours). Our framework is based on using any dynamic survival analysis model that supports competing risks in the form of estimating patient-level cumulative incidence functions. We consider three competing risks as to what happens first to a patient: awakening, being withdrawn from life-sustaining therapies (and thus deterministically dying), or dying (by other causes). For some patients, we do not know which of these happened first since they were still in a coma when data collection stopped (i.e., their outcome is censored). Competing risks models readily accommodate such patients. We demonstrate our framework by benchmarking three existing dynamic survival analysis models that support competing risks on a real dataset of 922 post-cardiac-arrest coma patients. Our main experimental findings are that: (1) the classical Fine and Gray model which only uses a patient's static features and summary statistics from the patient's latest hour's worth of EEG data is highly competitive, achieving accuracy scores as high as the recently developed Dynamic-DeepHit model that uses substantially more of the patient's EEG data; and (2) in an ablation study, we show that our choice of modeling three competing risks results in a model that is at least as accurate while learning more information than simpler models (using two competing risks or a standard survival analysis setup with no competing risks).
\end{abstract}

\setlength{\abovedisplayskip}{4pt plus 2pt}
\setlength{\belowdisplayskip}{4pt plus 2pt}
\setlength{\abovedisplayshortskip}{3pt plus 2pt}
\setlength{\belowdisplayshortskip}{3pt plus 2pt} 

\section{Introduction}
\label{sec:intro}

Cardiac arrest is one of the leading causes of death and disability worldwide, resulting in approximately 300,000 to 450,000 deaths annually in the U.S. alone (NIH\footnote{\url{https://www.nhlbi.nih.gov/health/cardiac-arrest}}, 2022) and a 43\% reported rate of suffering from cognitive impairment \citep{byron2021cognitive}. 
In this paper, we specifically focus on cardiac arrest patients who enter a coma and are admitted to the ICU, where they are placed on life-sustaining therapies (e.g., mechanical ventilation, cardiac support devices). Here, forecasting the neurological outcome of patients (i.e., the task of \emph{neurological prognostication}) is important: if physicians perceive a patient to have poor neurological prognosis, then they may discontinue life-sustaining therapies for the patient, which deterministically ends the patient's life.
Withdrawal of life-sustaining therapies accounts for 48\% of all nonsurvivors, with 31\% occurring in medically unstable patients and 17\% in medically stable patients \citep{matthews2017categorization}. This raises the possibility that some patients may have survived if different decisions had been made regarding their care. Neurological prognostication, therefore, is crucial in determining the appropriate treatment plan for each patient.

In recent years, a promising direction for neurological prognostication has been to take advantage of brain activity measurements using electroencephalography (EEG) \citep{abend2010use, Glass2013-ot, friberg2015survey}. A number of recent studies have demonstrated these EEG signals are predictive of patients' neurological outcomes (e.g., \citealt{thenayan2010electroencephalogram, rittenberger2012frequency, rossetti2012early, soholm2014prognostic, oh2015continuous, elmer2016clinically, westhall2016standardized, shekhar2023benefit}). In this paper, we use both EEG data recorded over time and some patient characteristics collected upon hospital admission.

While prognostication is essential, how it has been modeled in existing literature has some major limitations. First, binary classification has been the most common approach for modeling prognostication for post-cardiac arrest coma patients, where the two classes are poor neurological recovery or death (taken to be the ``positive'' class) and favorable neurological recovery with certain levels of consciousness at hospital discharge (the ``negative'' class) (e.g., \citealt{rossetti2016neurological, admiraal2021quantitative, moseby2020performance}). The goal is to achieve a high true positive rate (TPR) with a false positive rate (FPR) close to~0 (see, for instance, the overview by \citet{geocadin2019standards}). %
However, the existence of patients for whom life-sustaining therapies were withdrawn complicates this binary classification setup: including these particular patients in training data is problematic because we do not know what their neurological outcomes would have been if they had been kept on life-sustaining therapies (i.e., we do not know which of the two classes they belong~to). Some existing studies simply exclude such patients (e.g., \citealt{de2019predicting}). However, by excluding these patients, we ignore potentially useful information: these patients likely have characteristics that led to physicians withdrawing them from life-sustaining therapies. %

Another major limitation of existing work is ignoring the dynamic nature of both the EEG signals as well as physicians' decision-making and only focusing on using EEG data of a fixed period of time to make a prediction at a single point in time. For example, \citet{de2019predicting} only use EEG data between hours 34 to 36 after ICU admission to make a single prediction of the chance of favorable recovery for each patient; any patient with missing EEG data between hours 34 to 36 would be excluded, and EEG information outside of this time window would not be used by their prediction model. Meanwhile, \citet{admiraal2021quantitative} make a single prediction using EEG data 24 hours after cardiac arrest. In practice, physicians make decisions regarding treatment plans at different times after the patients have been admitted to the ICU (e.g., adjusting medications such as the dosage of vasopressors over time, or deciding whether to withdraw life-sustaining therapies), potentially using all information collected of the patient up until present time. To the best of our knowledge, no existing method has been developed for neurological prognostication of these coma patients that is truly \emph{dynamic}, where the training patients' time series can vary in length, and where we make predictions at any point in time.

Our main contribution in this paper is to propose a framework for neurological prognostication of post-cardiac-arrest coma patients that addresses both of the above major limitations. In particular, our framework is dynamic and directly models specific outcomes of interest as \emph{competing risks} in the sense that one outcome happening (e.g., withdrawal from life-sustaining therapies) prevents other outcomes from happening (e.g., awakening, dying of other causes). Moreover, our framework allows for outcomes to be \emph{censored}, meaning that for some patients, we do not get to see their eventual outcome since they were still in a coma by the time data collection stopped. That they were still alive at the end of data collection could be due to specific patient characteristics that make them more likely to be alive rather than to have been pulled off life-sustaining therapies or to have died of other causes.

Our framework builds on a class of existing dynamic survival analysis models that support competing risks; we refer to this class of models as \emph{dynamic competing risks (DCR) models} (we precisely define this class of models in Section~\ref{sec:background}). An example of such a model is Dynamic-DeepHit \citep{lee2019dynamic}. Roughly, a DCR model predicts, for any test patient with measurements up to time $t$, the probability of the patient experiencing each of the different competing events at any time after time $t$.

Even though we learn a DCR model with three competing risks for the neurological prognostication problem, at prediction time, one of these competing risks is often not of primary interest: whether a patient will be withdrawn from life-sustaining therapies. %
Even if we did predict who would be withdrawn from life-sustaining therapies, we would effectively be predicting how past decisions were made by physicians and \emph{not} what the true neurological outcomes of these patients would have been. Ideally, predictions of the latter should be what we use to assist with treatment decisions. For this reason, we show how to derive a binary classifier from the DCR model's predicted output that aims to predict whether a patient will awaken or die (of causes aside from withdrawal from life-sustaining therapies) within any user-specified time horizon (Section~\ref{sec:classifier}). Conceptually, whereas some existing work on neurological prognostication excluded patients who were withdrawn from life-sustaining therapy altogether from their analysis (e.g., \citealt{de2019predicting}), we are including such patients when training a DCR model, and only when using our classifier, we condition on (in a probabilistic sense) the test patient not being withdrawn from life-sustaining therapies in the future. Especially as this classifier is meant to help with decisions such as whether a patient should be withdrawn from life-sustaining therapies, a reasonable assumption is to condition on this event not happening yet. This binary classifier that we derive from the DCR model can classify variable-length input time series using any user-specified time horizon without needing to re-train the DCR model. We further develop a patient-specific heat map visualization for the classifier that is straightforward to interpret (Section~\ref{sec:visualization}).

In experiments on real data (cohort selection and other dataset details are in Section~\ref{sec:cohort}), we benchmark three DCR models to compare how accurate they are, and we also conduct an ablation study to show why our choice of modeling the competing risks setting with three competing risks is better than using two or one instead (Section~\ref{sec:experiment}). Note that the one competing risk setting reduces to a dynamic survival analysis setup without competing risks.

\subsection*{Generalizable Insights about Machine Learning in the Context of Healthcare}

For neurological prognostication of post-cardiac-arrest coma patients using EEG data, our paper is, to the best of our knowledge, the first to consider a dynamic problem setup. We believe that we have framed this prognostication problem in a manner that is more useful for clinical decision support compared to how it has been framed in existing literature.

As our dynamic problem setup and accompanying evaluation metrics have not previously appeared in literature for our specific clinical application, our two main experimental findings are novel: (1) in benchmarking three DCR models, we find that the classical competing risks model by \citet{fine1999proportional} that only uses static patient features and the summary information from the last hour of a patient's EEG data is highly competitive, achieving accuracy scores as high as the recently developed Dynamic-DeepHit model \citep{lee2019dynamic} that uses substantially more EEG data; and (2) our ablation study shows that our choice of modeling three competing risks results in a model that is at least as accurate while providing more information than a model that uses two competing risks or a dynamic survival analysis model without competing risks. These findings suggest that researchers working on the same clinical application may want to also consider modeling at least the three competing risks we consider (or even finer-grain versions of some of them, such as accounting for more causes of death), and also trying the classical Fine and Gray model as a baseline.

From a technical standpoint, our paper does not introduce a new model. Instead, our paper demonstrates how to effectively use any existing DCR model to address a specific clinical problem. The novelty is thus in the application of existing DCR models and also in our proposal of a classifier (derived from a DCR model) and an accompanying heat map visualization for this classifier. The crucial insight of the classifier that we develop is to condition on a particular event---an action taken by a physician that inevitably ends the patient's life---\emph{not} happening in the future of the patient because the output of the classifier is meant to help with deciding on whether to take this action (where a reasonable assumption is that we have not taken the action as doing so has a permanent consequence). We suspect that this same idea would be relevant in various other clinical problems.

\section{Background}\label{sec:lr}
\label{sec:background}

Our paper builds on a specific class of existing dynamic survival analysis models, which we refer to as dynamic competing risks (DCR) models. We review this class of models in Section~\ref{sec:problem-setup}, where we also state the dynamic problem setting that our framework uses. We briefly give a concrete example of a DCR model (Dynamic-DeepHit by \citet{lee2019dynamic}) in Section~\ref{sec:dynamic-deephit}. Note that throughout this paper, for any positive integer $m$, we frequently use the notation $[m]:=\{1,2,\dots,m\}$. We typically use uppercase variables to refer to random variables whereas lowercase variables refer to constants, realized values of random variables, or dummy indices (e.g., training data indices).

\subsection{Dynamic Problem Setup and DCR Models}
\label{sec:problem-setup}

\paragraph{Training data}
We assume that we have a training dataset consisting of $n$ patients. For each training patient~$i\in[n]$, we observe a times series with a total of $L_i$ time steps, where at each time step, we observe $d$ features. Specifically, we observe the feature vectors $X_i^{(1)},X_i^{(2)},\dots,X_i^{(L_i)}\in\mathbb{R}^d$, where $X_i^{(\ell)}\in\mathbb{R}^d$ is the $i$-th patient's feature vector at time step $\ell\in[L_i]$ (time steps are sorted chronologically, so time step $L_i$ is the last time step observed for training patient~$i$). Moreover, time step $\ell\in[L_i]$ happens at a time that is recorded as a real number~$T_i^{(\ell)}\in\mathbb{R}$, meaning that the amount of time that elapses between time steps~$\ell$ and~$\ell+1$ is $T_i^{(\ell+1)}-T_i^{(\ell)}$. A common assumption is to set the initial time step's time to be $T_i^{(1)}=0$. Note that for the $d$ features that are tracked over time, it is possible that some always stay the same (e.g., age upon hospital admission). For ease of exposition, we do not introduce additional notation that separates static from time-varying features although separating these two types of features could be done in practice.

As the above notation suggests, patients' time series can vary in length (e.g., one patient could have EEG data recorded every second for~6 hours, whereas another could have EEG data recorded every second for~12 hours). For the real data we consider later, the time series are regularly sampled (i.e., the amount of time that elapses between consecutive time steps is the same) but in general, DCR models can handle irregularly sampled time series.

In terms of ground truth information, for each training patient $i\in[n]$, we assume that we observe two quantities:
\begin{itemize}[itemsep=0pt,parsep=0pt,topsep=2pt] %
\item (Event indicator) We observe which of $k$ different competing events happened first to the $i$-th patient, or alternatively we could also observe that none of the competing events happened by the time training data collection stopped. This information is stored in the event indicator $K_i\in\{0,1,2,\dots,k\}$. For example, in the neurological prognostication problem, we have $k=3$ and the competing events are awakening ($K_i=1$), dying of causes aside from withdrawal from life-sustaining therapies ($K_i=2$), and withdrawal from life-sustaining therapies ($K_i=3$). The special value of $K_i=0$ means that by the time data collection stopped, none of the competing events happened.
\item (Event time) We also observe the time $Y_i\in\mathbb{R}$ for when the first competing event happened or, if none of them happened, then $Y_i$ is the time when data collection stopped for the $i$-th patient (i.e., the time of ``censoring''). Note that at any time step $\ell\in[L_i]$, the time until the first competing event or censoring happens is $Y_i - T_i^{(\ell)}$. %
\end{itemize}
In summary, for training patient $i\in[n]$, we observe an event indicator $K_i\in\{0,1,\dots,k\}$, an event time $T_i\in\mathbb{R}$, and an input time series of feature vectors $X_i^{(1)},X_i^{(2)},\dots,X_i^{(L_i)}\in\mathbb{R}^d$ at corresponding times $T_i^{(1)},T_i^{(2)},\dots,T_i^{(L_i)}\in\mathbb{R}$. As shorthand notation for referring to the entire observed time series, we write $Z_i := \big((X_i^{(1)},X_i^{(2)},\dots,X_i^{(L_i)}), (T_i^{(1)},T_i^{(2)},\dots,T_i^{(L_i)})\big)$. %

We model the training data $(Z_1,K_1,Y_1),\dots,(Z_n,K_n,Y_n)$ to be i.i.d. To formally state how each training point is generated, we define a few probability distributions: $\mathbb{Q}_{\mathcal{T}}$ denotes an underlying probability distribution over variable-length time series, $\mathbb{Q}_{\mathcal{E}}(Z)$ denotes an underlying conditional probability distribution over nonnegative time durations across all $k$ competing events given a specific time series $Z$ (i.e., a random sample from $\mathbb{Q}_{\mathcal{E}}(Z)$ yields a vector in $\mathbb{R}^k$ where the $j$-th entry of the vector is a random time duration until competing event $j\in[k]$ happens), and $\mathbb{Q}_{\mathcal{C}}(Z)$ denotes the underlying conditional probability distribution over nonnegative time durations until censoring. %
Specifically, the $i$-th training point $(Z_i,K_i,Y_i)$ is generated as follows:
\begin{enumerate}[itemsep=0pt,parsep=0pt,topsep=2pt] %
\item We sample time series $Z_i$ (with $L_i$ time steps and last time $T_i^{(L_i)}$ using our earlier notation) from $\mathbb{Q}_{\mathcal{T}}$.
\item We sample the true nonnegative time durations $(\Xi_{i,1},\Xi_{i,2},\dots,\Xi_{i,k})$ from $\mathcal{Q}_{\mathcal{E}}(Z_i)$ (so that $\Xi_{i,1}$ is the time until the 1st competing event happens, $\Xi_{i,2}$ is the time until the 2nd competing event happens, etc).
\item We sample the true nonnegative time duration $\Xi_{i,0}$ (time until censoring) from a conditional distribution $\mathbb{Q}_{\mathcal{C}}(Z_i)$.
\item We set $K_i := \arg\min_{j=0,1,\dots,k} \Xi_{i,j}$, and $Y_i := T_i^{(L_i)} + \Xi_{i,K_i}$.
\end{enumerate}
Note that the competing events are ``exhaustive'' in the sense that with probability 1, either one of them happens or censoring happens.

\vspace{-.2em}
\paragraph{Prediction target}

We model a test patient's data using the same distributions that we introduced for training data. However, for the test patient, our goal will never be to predict whether the test patient is censored. In particular, we model a test patient with time series~$Z$, event indicator $K$ (always a value in $\{1,\dots,k\}$), and event time $Y$ as follows:
\begin{enumerate}[itemsep=0pt,parsep=0pt,topsep=2pt] %
\item We sample time series $Z=\big((X^{(1)},\dots,X^{(L)}),(T^{(1)},\dots,T^{(L)})\big)$ (using notation similar to that of training data) from $\mathbb{Q}_{\mathcal{T}}$. Note that $Z$ has $L$ time steps.
\item We sample the true nonnegative time durations $(\Xi_1,\Xi_2,\dots,\Xi_k)$ from $\mathbb{Q}_{\mathcal{E}}(Z)$.
\item We set $K := \arg\min_{j=1,\dots,k} \Xi_j$, and $Y := T^{(L)} + \Xi_K$.
\end{enumerate}
Even though time series $Z$ is generated so that it has $L$ time steps, in how we set up the prediction task next, we do not observe all time steps immediately. Instead, we progressively observe more of $Z$ over time, similar to what would happen in a real clinical context. In particular, we state our prediction task to depend on time $t\in\mathbb{R}$ and use the random variable $Z^{(\le t)}$ to denote time series $Z$ limited to information up until time $t$. Specifically, we aim to predict the so-called \emph{cumulative incidence function} (CIF) of event $j\in[k]$, which is the probability of event $j$ happening within time duration $\Delta\ge0$ starting from time $t\in\mathbb{R}$, given a time series observed up until time $t$. Formally, we write the CIF as
\begin{equation}
F_j(\Delta \mid z,t)
:= \mathbb{P}(Y \le t + \Delta, K = j \mid Z^{(\le t)} = z^{(\le t)}, Y > t)\qquad\text{for }\Delta\ge 0,
\label{eq:CIF}
\end{equation}
where $t$ is the time that we are making a prediction at, and $z$ is any specific realization of random variable $Z$ (again, the superscript ``$^{(\le t)}$'' restricts time to be up until $t$).

Note that we have intentionally stated the CIF in the dynamic setting with variable-length time series, where we can make predictions at different points in time. The classical version of the CIF \citep{gray1988class,fine1999proportional} is stated in the ``static'' setting without time series and can be viewed as a special case of the dynamic setup we have described, where all time series sampled from $\mathbb{Q}_{\mathcal{T}}$ have exactly one time step, the recorded time of this first time step is always just taken to be 0, and we only ever evaluate equation~\eqref{eq:CIF} at $t=0$. Separately, if the number of competing risks is equal to $k=1$, then the entire setup we have described would instead be for dynamic survival analysis without competing risks. In fact, one could show that the static setting with one competing risk simply reduces to the classical right-censored survival analysis setup (e.g., see the random censoring setup described in Section 3.2 of the textbook by \citet{kalbfleisch2002statistical}).

\vspace{-.2em}
\paragraph{Dynamic competing risks (DCR) models}
The class of DCR models that our framework for neurological prognostication builds on is any model that can predict CIFs as given in equation~\eqref{eq:CIF}. For example, Dynamic-DeepHit \citep{lee2019dynamic} and SurvLatent ODE \citep{moon2022survlatent} are DCR models. Note that any classical competing risks model (e.g., \citealt{fine1999proportional}) that does not actually handle variable-length time series could be made into a DCR model in a simple manner: simply only use the last time step's feature vector to predict. Some existing dynamic survival analysis models (such as DDRSA \citep{venkata2022intervene}) that were not originally developed to support competing events could be modified to estimate CIFs as well. To give a sense of how a DCR model works, we review Dynamic-DeepHit next. Note that we specifically review a DCR model that directly models variable-length time series without manual feature engineering or resorting to, for instance, only ever using the last time step of an input time series.

\subsection{Example of a DCR Model: Dynamic-DeepHit}
\label{sec:dynamic-deephit}

\newcommand{\pmfTimeIndex}{u}
\newcommand{\pmfTimeIndexDummy}{v}

We provide an overview of Dynamic-DeepHit \citep{lee2019dynamic}, deferring details to the original paper.\footnote{Note that \citet{lee2019dynamic}~explicitly keep track of a separate vector per time step indicating which of the $d$ features are missing. Instead of introducing notation for such a ``missingness'' boolean vector, we can augment our original feature vector to include such missingness indicator variables.} Importantly,
Dynamic-DeepHit discretizes possible values for time duration $\Delta$ in equation~\eqref{eq:CIF} into $m$ unique values ${\Delta_1<\Delta_2<\cdots<\Delta_m}$. We assume that $\Delta_1 > 0$ and that $\Delta_m$ is an upper bound on possible durations encountered across all events $j\in[k]$ (i.e., all $k$ events happen within time duration~$\Delta_m$). In what follows, we regularly use the variables $u,v\in[m]$ to denote indices of the discretized values of $\Delta$. Dynamic-DeepHit estimates a probability mass function variant of the CIF for event $j\in[k]$ given by
\begin{equation}
O_j(\Delta_{\pmfTimeIndex} \mid z, t) := \mathbb{P}( \Delta_{\pmfTimeIndex - 1} < Y - t \le \Delta_{\pmfTimeIndex}, K = j \mid Z^{(\le t)} = z^{(\le t)}, Y > t)
\quad\text{for }\pmfTimeIndex\in[m],
\label{eq:CIF-pmf}
\end{equation}
where we define $\Delta_0:= 0$ to handle the case when we plug in $u=1$. Before explaining why the above function behaves like a probability mass function, we point out that one can readily verify that the CIF for event $j$ from equation~\eqref{eq:CIF} satisfies the equality
\begin{equation}
F_j(\Delta_{\pmfTimeIndex} \mid z, t)
= \sum_{\pmfTimeIndexDummy=1}^\pmfTimeIndex O_j(\Delta_{\pmfTimeIndexDummy} \mid z, t)
\qquad\text{for }\pmfTimeIndex\in[m].
\label{eq:CIF-discrete}
\end{equation}
Thus, so long as we can estimate $O_j(\Delta_{\pmfTimeIndex} \mid z, t)$ in equation~\eqref{eq:CIF-pmf}, then we obtain an estimate of the CIF in equation~\eqref{eq:CIF} albeit only along a discrete time grid $\Delta\in\{\Delta_1,\dots,\Delta_m\}$.

As for why $O_j(\Delta_{\pmfTimeIndex} \mid z, t)$ behaves like a probability mass function, note that
\begin{equation}
\sum_{j=1}^k \sum_{\pmfTimeIndex=1}^m O_j(\Delta_{\pmfTimeIndex} \mid z, t)
= \sum_{j=1}^k F_j(\Delta_m \mid z, t) = 1,
\label{eq:sum-to-1-constraint}
\end{equation}
where we have used equation~\eqref{eq:CIF-discrete} and the assumption that $\Delta_m$ is chosen as an upper bound on possible durations across all $k$ competing events.

\begin{figure}[t]
\centering\vspace{-.1em}
\includegraphics[width=0.99\linewidth]{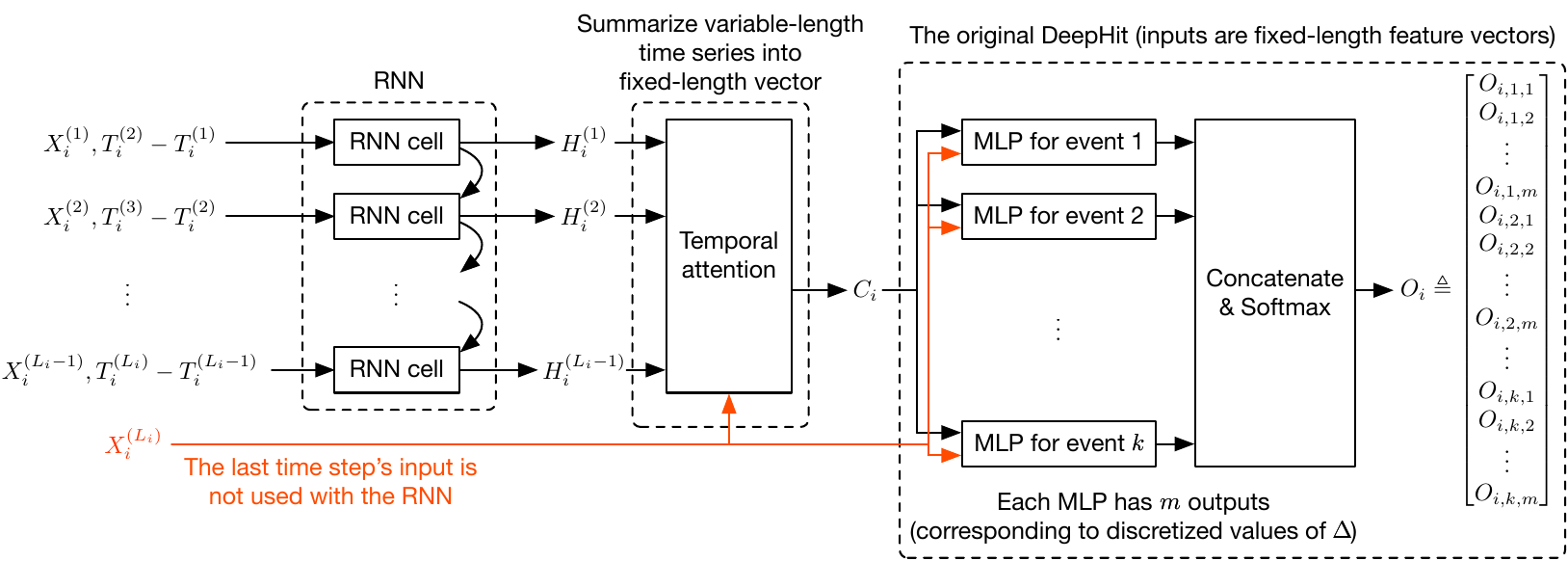}\vspace{-1.1em}
\caption{Dynamic-DeepHit network architecture.}
\label{fig:ddh-model}
\vspace{-1em}
\end{figure}

With this motivation, Dynamic-DeepHit models $O_j(\Delta_{\pmfTimeIndex}\mid z,t)$ in equation~\eqref{eq:CIF-pmf} using a neural network. For the $i$-th training time series $Z_i:=\big((X_i^{(1)},\dots,X_i^{(L_i)}),(T_i^{(1)},\dots,T_i^{(L_i)})\big)$, we specifically estimate $O_j(\Delta_{\pmfTimeIndex}\mid Z_i, T_i^{(L_i)})$ to be equal to $O_{i,j,\pmfTimeIndex}$ (with $i\in[n],j\in[k],u\in[m]$), where $O_{i,j,\pmfTimeIndex}$ is shown on the right side of Figure~\ref{fig:ddh-model}; we collect all the $O_{i,j,u}$ values specific to the $i$-th patient in the vector $O_i\in\mathbb{R}^{k \cdot m}$. We compute $O_i$ from $Z_i$ as follows:
\begin{enumerate}[itemsep=0pt,parsep=0pt,topsep=2pt]
\item We first feed the input time series $Z_i$ into a user-specified RNN (with $p$ output features per time step, where the choice of $p$ is up to the user), where we slightly transform what the input looks like per time step. Specifically at time step $\ell\in[L_i - 1]$, the input to the RNN is taken to be $(X_i^{(\ell)}, T_i^{(\ell+1)}-T_i^{(\ell)})$, i.e., we provide both a feature vector and a time duration to get to the next time step. However, the last time step's feature vector $X_i^{(L_i)}$ is not used with the RNN (but will be used later). The RNN's output at time step $\ell\in[L_i-1]$ is denoted as $H_i^{(\ell)}\in\mathbb{R}^p$. This first step is shown on the left side of Figure~\ref{fig:ddh-model}.
\item The second step is uses the variable-length time series $(H_i^{(1)},\dots,H_i^{(L_i-1)})$ and the last time step's feature vector $X_i^{(L_i)}$ to compute a fixed-length summary vector $C_i\in\mathbb{R}^p$. To do this, we set $C_i = \sum_{\ell=1}^{L_i-1} a_{\ell} H_i^{(\ell)}$, where
\[
\begin{bmatrix}
a_1 \\
a_2 \\
\vdots \\
a_{L_i-1}
\end{bmatrix}
:= \text{softmax}\!
   \left(
   \begin{bmatrix}
   f_{\text{attention}}((H_i^{(1)}, X_i^{(L_i)})) \\
   f_{\text{attention}}((H_i^{(2)}, X_i^{(L_i)})) \\
   \vdots \\
   f_{\text{attention}}((H_i^{(L_i-1)}, X_i^{(L_i)}))
\end{bmatrix}
   \right) \in [0,1]^{L_i-1},
\]
and $f_{\text{attention}}$ is a user-specified feed-forward neural network, such as a multilayer perceptron (MLP), that outputs a single real number. This second step is shown in the middle of Figure~\ref{fig:ddh-model}.
\item We then combine vector $C_i$ and the last time step's feature vector $X_i^{(L_i)}$ as the input to $k$ different MLPs (one per competing risk) that each outputs $m$ numbers, and the overall concatenated output is passed through a softmax layer to produce the final output $O_i$ (the softmax enforces the constraint in equation~\eqref{eq:sum-to-1-constraint}). Note that this third step corresponds to the original DeepHit model \citep{lee2018deephit} that is meant for handling input data that are fixed-length feature vectors rather than variable-length time series. This step is shown on the right side of Figure~\ref{fig:ddh-model}.
\end{enumerate}
There is one last neural network component that is not shown in Figure~\ref{fig:ddh-model} as it is not used to compute the output vector $O_i$: Dynamic-DeepHit also requires that at time step $\ell\in[L_i-1]$, the RNN on the left side of Figure~\ref{fig:ddh-model} can output an estimate $\widehat{X}_i^{(\ell+1)}$ of the next time step's feature vector $X_i^{(\ell+1)}$. There are different ways to achieve this. For example, at time step $\ell\in[L_i-1]$, we can feed $H_i^{(\ell)}$ (along with the time duration to get to the next time step) into a user-specified MLP $f_{\text{next-time-step}}$ with $d$ output features to produce the estimate $\widehat{X}_i^{(\ell+1)}$. Alternatively, for the RNN in Figure~\ref{fig:ddh-model}, we could choose it to be a type of RNN that already distinguishes between hidden state vectors and output state vectors (e.g., LSTMs \citep{hochreiter1997long}), in which case we let the hidden state vectors be what we denoted as the $H_i^{(\ell)}$ variables, and we use the output state vectors to predict the next steps' feature vectors (the output state vector would need to consist of $d$ entries).

\vspace{-.2em}
\paragraph{Training}
The final loss used to train a Dynamic-DeepHit model is the sum of three terms, two of which make up the original DeepHit loss (a negative log likelihood term and a ranking loss term), and the last term asks that each next feature vector estimate $\widehat{X}_i^{(\ell+1)}$ is close to $X_i^{(\ell+1)}$ (using, for instance, squared Euclidean distance).

\vspace{-.2em}
\paragraph{Prediction}
At test time, we could feed any observed test time series (of arbitrarily nonzero length) as input to the neural network in Figure~\ref{fig:ddh-model} to produce an estimate of the probability mass function of the CIF in equation~\eqref{eq:CIF-pmf} that we can then use to estimate the CIF with using equation~\eqref{eq:CIF-discrete}. We could trivially accommodate the setting where we see more of a test time series over time since the neural network accepts a variable-length input time series.

\section{Framework for Neurological Prognostication}\label{sec:methodology}

Any DCR model could be applied to the problem of neurological prognostication for post-cardiac-arrest coma patients. As stated in Section~\ref{sec:background}, we can take the number of competing risks to be $k=3$ corresponding to awakening $(K=1)$, dying (not of withdrawal from life-sustaining therapies) $(K=2)$, or withdrawal from life-sustaining therapies (and thus dying as a result) $(K=3)$. A key goal of our framework is to help clinicians interpret the information contained in the CIFs (equation~\eqref{eq:CIF}) predicted by a DCR model.

Using the three competing risks stated above, we derive a binary probabilistic classifier from an already trained DCR model (Section~\ref{sec:classifier}). The resulting classifier can then be used to produce a patient-specific prediction heat map visualization that aims to be straightforward for a clinician to interpret (Section~\ref{sec:visualization}). Separately, standard binary classification evaluation metrics could be used for the derived classifier, which supplement survival analysis evaluation metrics that already exist for DCR models.

\subsection{A Derived Binary Classifier}
\label{sec:classifier}

As discussed in Section~\ref{sec:intro}, predicting whether a patient will be withdrawn from life-sustaining therapies is often not of primary interest since this is a human-made decision that deterministically ends a patient's life, and we do not actually know for sure whether the patient would have instead awakened or died of other causes in the ICU. For any test patient's time series up to time~$t$, we now derive a binary probabilistic classifier that conditions on the event that \emph{the test patient is never withdrawn from life-sustaining therapies} (we refer to this event as the ``non-withdrawal event''). As we aim to develop a decision support tool to help physicians decide on whether to withdraw life-sustaining therapies, a reasonable assumption is that the patients are kept on these therapies, especially since withdrawal of these therapies has a permanent effect (in Section~\ref{sec:discussion}, we comment on whether this conditioning makes sense).

After conditioning on the non-withdrawal event, we then compute the probabilities of the remaining two competing events happening within a time duration $\Delta>0$.
This conditional probability can be computed as follows, reusing notation from Section~\ref{sec:background}:
\begin{align}
P_{\text{awaken}}(\Delta \mid z, t)
& := \mathbb{P}(Y\le t+\Delta,\overbrace{K=1}^{\text{awaken}}~~\mid \!\!\overbrace{K\in\{1,2\}}^{\substack{\text{awaken or death (not}\\\text{by withdrawal from}\\\text{life-sustaining therapies)}}}\!\!,Z^{(\le t)}=z^{(\le t)},Y>t) \nonumber\\
& =\frac{\mathbb{P}(Y\le t+\Delta,K=1\mid Z^{(\le t)}=z^{(\le t)},Y>t)}{\mathbb{P}(K\in\{1,2\}\mid Z^{(\le t)}=z^{(\le t)},Y>t)} \nonumber\\
& =\frac{\mathbb{P}(Y\le t+\Delta,K=1\mid Z^{(\le t)}=z^{(\le t)},Y>t)}{\mathbb{P}(K=1\mid Z^{(\le t)}=z^{(\le t)},Y>t)+\mathbb{P}(K=2\mid Z^{(\le t)}=z^{(\le t)},Y>t)} \nonumber\\
& =\frac{F_1(\Delta\mid z,t)}{F_1(\infty\mid z,t)+F_2(\infty\mid z,t)}.
\label{eq:prob-awaken-derivation}
\end{align}
Thus, if we have trained a DCR model that has an estimate $\widehat{F}_j(\Delta\mid z,t)$ for each CIF $F_j(\Delta\mid z,t)$, then we can directly plug in these CIF estimates into the right-hand side above to yield the estimated conditional probability
\begin{equation}\label{eq:prob-awake}
\widehat{P}_{\text{awaken}}(\Delta \mid z, t)
:=
  \frac{\widehat{F}_1(\Delta\mid z,t)}
       {\widehat{F}_1(\infty\mid z,t)
        +\widehat{F}_2(\infty\mid z,t)}.
\end{equation}
We could similarly estimate the probability for death (not by withdrawal from life-sustaining therapies) by
\begin{equation}\label{eq:prob-death}
\widehat{P}_{\text{death (not withdrawal)}}(\Delta \mid z, t)
:=
  \frac{\widehat{F}_2(\Delta\mid z,t)}
       {\widehat{F}_1(\infty\mid z,t)
        +\widehat{F}_2(\infty\mid z,t)}.
\end{equation}
We thus have a binary probabilistic classifier between the two classes ``awaken'' and ``death (not by withdrawal from life-sustaining therapies)'' defined for a specific time $t$ and time duration $\Delta$: if the ratio $\frac{\widehat{P}_{\text{death (not withdrawal)}}(\Delta \mid z, t)}{\widehat{P}_{\text{awaken}}(\Delta \mid z, t)}$ is below a threshold value of~1, then we predict ``awaken''. The threshold of 1 could of course be tuned (e.g., to achieve some desired tradeoff between TPR and FPR on some validation set). Note that it only makes sense to plug in times $t$ that are at least the earliest time encountered in time series $z$.%

Importantly, the binary classifier we just described was derived using estimated CIFs. Existing DCR models like Dynamic-DeepHit estimate CIFs in a manner that would include patients of all three competing risks as well as those who were censored. In particular, we do not have to, for example, exclude patients who are censored or who were withdrawn from life-sustaining therapies from the analysis. Moreover, this classifier can be constructed for any time $t$ after the earliest time in the observed test time series $Z=z$ and for any choice of user-specified time duration $\Delta>0$, without any re-training of the underlying DCR model.

\vspace{-.2em}
\paragraph{Technical remark}
The estimated probabilities $\widehat{P}_{\text{awaken}}(\Delta \mid z, t)$ in equation~\eqref{eq:prob-awake} and ${\widehat{P}_{\text{death (not withdrawal)}}(\Delta \mid z, t)}$ in equation~\eqref{eq:prob-death} do not, in general, sum to~1. This is intentional. Again, each probability is derived based on equation~\eqref{eq:prob-awaken-derivation}, where the final denominator is the probability that a patient with time series $z$ up to time $t$ experiences an eventual outcome that is either $K=1$ or $K=2$. From an interpretation standpoint, an appealing aspect of how $\widehat{P}_{\text{awaken}}(\Delta \mid z, t)$ (and similarly ${\widehat{P}_{\text{death (not withdrawal)}}(\Delta \mid z, t)}$) is defined is as follows. For a fixed time~$t$, consider two time durations $\Delta$ and $\Delta'$, where $\Delta<\Delta'$ (for example, $\Delta=24$ hours and $\Delta'=48$ hours). Then intuitively it makes sense that the probability of someone awakening within time duration~$\Delta'$ should be larger than the probability of someone awakening within the time duration~$\Delta$, since $\Delta'>\Delta$. In other words, we would like $\widehat{P}_{\text{awaken}}(\Delta \mid z, t) \le \widehat{P}_{\text{awaken}}(\Delta' \mid z, t)$. This property indeed holds for how we have defined equations~\eqref{eq:prob-awake} (and a similar result holds for $\widehat{P}_{\text{death (not withdrawal)}}$). This property would \emph{not} be guaranteed to hold if instead we had changed the denominators of equations~\eqref{eq:prob-awake} and~\eqref{eq:prob-death} to $\widehat{F}_1(\Delta\mid z,t)+\widehat{F}_2(\Delta\mid z,t)$, which would ensure that the probabilities sum to~1.

\subsection{Patient-Specific Heat Map Visualization}
\label{sec:visualization}

\begin{figure}[t]
  \centering
  \subfigure[Example Patient 1: died (not from withdrawal of life-sustaining therapies) at hour 17]{\includegraphics[width=0.99\linewidth]{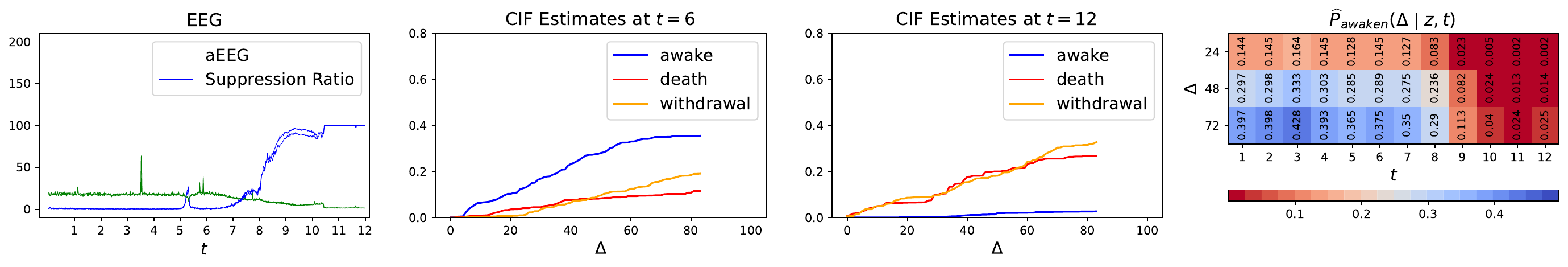}\label{fig:p1-visual}}
  \vfill
  \subfigure[Example Patient 2: was still in a coma at hour 118]{\includegraphics[width=0.99\linewidth]{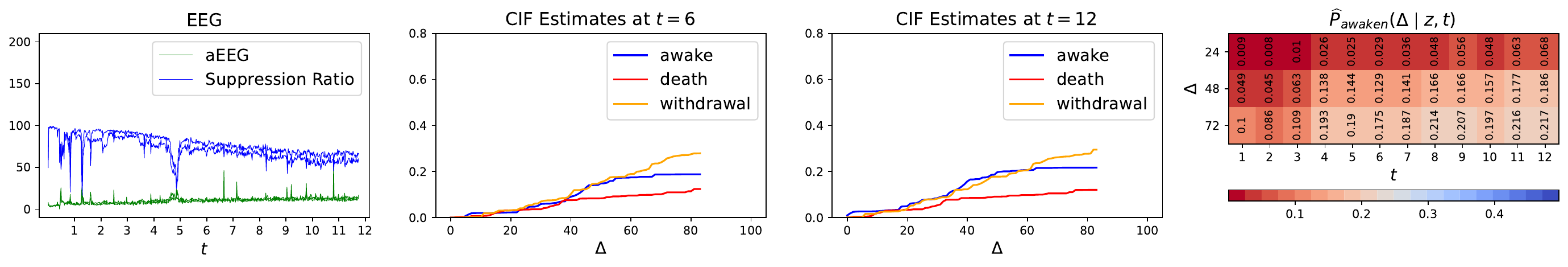}\label{fig:p2-visual}}
  \vfill
  \subfigure[Example Patient 3: awakened at hour 54]{\includegraphics[width=0.99\linewidth]{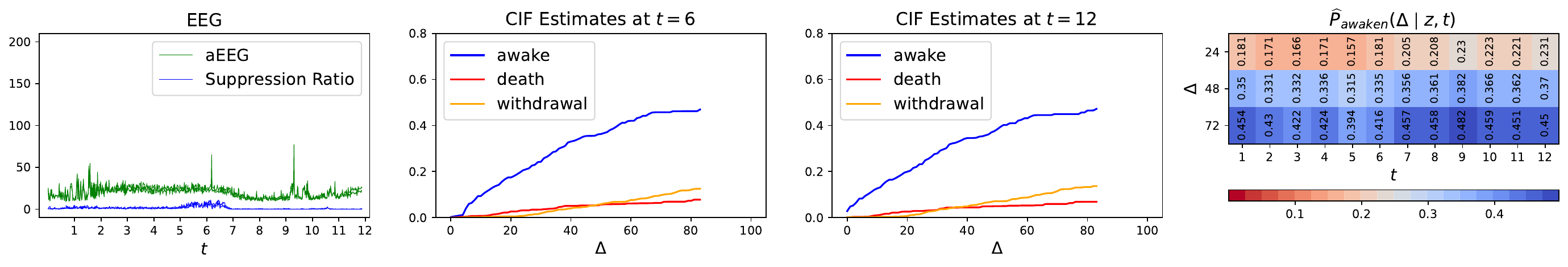}\label{fig:p5-visual}}
  \vfill
  \subfigure[Example Patient 4: died (not from withdrawal of life-sustaining therapies) at hour 71]{\includegraphics[width=0.99\linewidth]{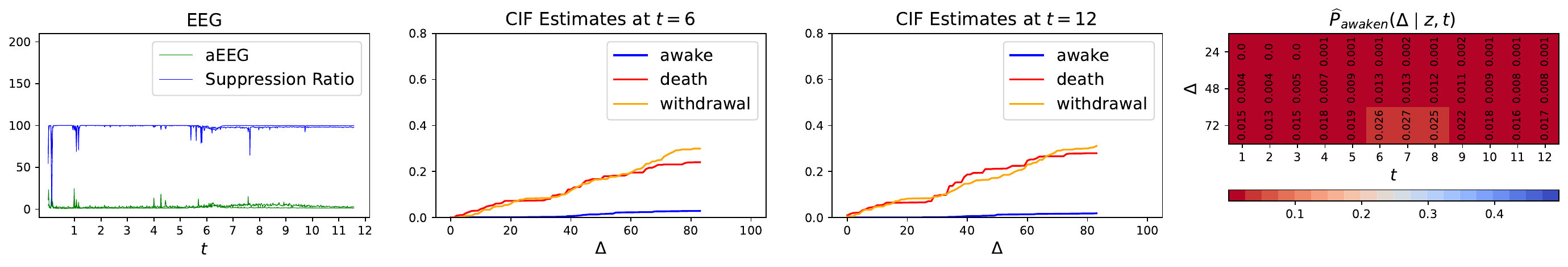}\label{fig:p4-visual}}\vspace{-.75em}
  \caption{For each example patient (panels $(a)$-$(d)$), we show time series of two summary EEG features \emph{aEEG} and \emph{suppression ratio} (first column plot), estimated CIFs at hours $t=6$ (second column plot) and $t=12$ (third column plot), and our proposed heat map visualization (fourth column plot). Note that aEEG values for normal brain activity should be within a certain range (constantly being lower than 5 or higher than 25 is usually considered abnormal), and higher suppression ratio values are a sign of more severe dysfunction or injury.} %
  \label{fig:cif-eeg}
  \vspace{-2em}
\end{figure}

We propose a heat map visualization specific to any patient that shows how the predicted probability of awakening ($\widehat{P}_{\text{awaken}}(\Delta \mid z, t)$ in equation~\eqref{eq:prob-awake}) changes for the patient as we observe more of the patient's time series, as shown in the fourth column plot of each row in Figure~\ref{fig:cif-eeg}. The experimental setup that led to this figure is explained in more detail in Section~\ref{sec:experiment}. For now, focusing on any one of the fourth-column heat maps of Figure~\ref{fig:cif-eeg}, we have the
horizontal axis correspond to the prediction time $t$ while the vertical axis is the time duration $\Delta$. We specifically choose the $\Delta$'s to be equivalent of the next 24 to 72 hours, which is of interest according to clinician feedback we have received. Using Patient 1 in Figure~\ref{fig:cif-eeg} as an example, we observe a drastic decrease in the probability of awakening starting at $t=9$ across all $\Delta$ values evaluated. %

\section{Cohort}\label{sec:cohort}
We examine proprietary hospital data collected in a single medical center from 2010 to 2019 (the specific medical center has been blinded for reviewing purposes). %
The dataset includes
patients who suffered sudden cardiac arrest, were successfully resuscitated, and survived to hospital admission at the medical center. EEG is initiated and monitored for these patients continuously as a routine standard of care for several days after cardiac arrest. %
As stated in the previous sections, we focus on three outcomes: awakening from the coma, %
dying (not from withdrawal of life-sustaining therapies, such as from brain death or rearrest) %
and withdrawal from life-sustaining therapies (due to perceived poor neurological prognosis, leading to death). %
We remark that there is a fourth outcome that is possible but that we exclude from analysis: patients could be withdrawn from life-sustaining therapies for non-neurological reasons such as a do-not-resuscitate order. As our focus is on neurological prognostication, we ignore this outcome in which a decision is made disregarding neurological prognosis. Note that in this study, we only care about the first occurrence of the event that ceases the coma status of a patient, i.e. if a patient awakened at some point and then still died shortly afterward, we would only consider awakening as the outcome of the patient. %

\vspace{-.2em}
\paragraph{Dataset characteristics}
After excluding patients whose cause of death is withdrawal for non-neurological reasons, %
we have a dataset consisting of 922 patients. For the 922 patients, summary statistics of their characteristics and the time-to-event are shown in an appendix (Table \ref{tab:summary-stats}). Out of the 922 patients, 271 (29.4\%) of them awakened %
at some point, 189 (20.5\%) of them died
(not from withdrawal from life-sustaining therapies), %
432 (49.6\%) of them died from withdrawal of life-sustaining therapies due to poor perceived neurological prognosis, and 30 (3.3\%) of them were still in a coma when data collection stopped. %

\vspace{-.2em}
\paragraph{EEG time series data} %
Raw EEG waveform data, typically recorded at 256Hz from 22 electrodes distributed on the brain according to standard clinical practice, are processed using FDA-approved clinical software (\url{https://www.persyst.com/}) to quantify more than 2,500 clinically-understood features every second (e.g., amplitude, frequency decompositions, suppression ratio, etc). %
For ease of exposition, here we focus on 12 features (see Appendix~\ref{sec：app-data} for details) that have been proven to be useful in this domain: \textit{suppression ratio} and \textit{amplitude-integrated EEG (aEEG)} \citep{oh2015continuous, elmer2016group}. We preprocess and downsample the raw second-by-second data to have one measurement (per feature) per hour in two steps: (1) For each of the 12 EEG features used in this study, we downsample the data by taking the average value of each consecutive non-overlapping block of 60 seconds to get a single value (i.e., we first downsample the data so that each time step corresponds to one minute). (2) We then further downsample the minute-resolution EEG signals so that each time step corresponds to an hour, and for each of the 12 EEG features, we take 6 summary statistics (minimum, maximum, mean, 25\% percentile, median, and 75\% percentile) as the final features to be used (i.e., for each time step corresponding to 1 hour, we end up with a total of $12\times6 = 72$ summary features). %
We remark that the data after downsampling still captures most of the variation from the raw data and does not have much impact on prediction accuracy. Downsampling is mainly to reduce computation time.

Note that in how we curated the data, we only use at most 12 hours of EEG data per patient (so that if a patient has more than 12 hours worth of EEG data, we ignore the EEG data after 12 hours). This is a limitation of the dataset curation process that could be changed in future work. From a modeling perspective, DCR models could in principal use arbitrarily long EEG time series, subject to hardware memory constraints in practice.

\vspace{-.2em}
\paragraph{Static features} Upon hospital admission, a number of features are collected for the patient that we treat as static features, such as demographic information (e.g. age, gender), characteristics of the patient's initial cardiac arrest collapse (e.g., arrest location, initial arrest rhythm, category of the cardiac arrest), initial coma status, and medical history. A full list of the 43 static features we use can be found in Appendix~\ref{sec：app-data}. %

\smallskip
\noindent
In summary, accounting for both the time-varying EEG features and the static features, we use a total of $d=72+43=115$ final features per time step when we train a DCR model.

\section{Experiments} \label{sec:experiment}
In this section, we run experiments on the dataset described in the previous section using three DCR models: a classical competing risks model by \citet{fine1999proportional} using only the last observed time step of each input time series (details are in Appendix~\ref{sec:app-fine-and-gray}), Dynamic-DeepHit \citep{lee2019dynamic} and DDRSA \citep{venkata2022intervene}. Also, note that the original DDRSA model by \citet{venkata2022intervene} does not support competing risks and we modify its network structure to accommodate competing risks (details are in Appendix~\ref{sec:app-ddrsa-competing-risk}). For simplicity, we refer to the competing-risk-adapted version of DDRSA as DDRSA despite the modification we make. We specifically aim to:
\begin{itemize}[itemsep=0pt,parsep=0pt,topsep=2pt]
\item (Section~\ref{sec:accuracy}) examine how accurate these models using the standard survival analysis metric of concordance index (abbreviated c-index; this is a value between 0 and 1 where higher is better) \citep{harrell1982evaluating} as well as AUROC of binary classifiers derived using our approach in Section~\ref{sec:classifier},
\item (Section~\ref{sec:multiple-visualizations}) provide examples of patient-specific visualizations (time series of two summary EEG signals, estimated CIFs, and the heat map visualization we described in Section~\ref{sec:visualization})
\item (Section~\ref{sec:ablation}) show that using a setup with fewer than the three competing risks we consider result in models that are not as good.
\end{itemize}
\vspace{-.5em}
\paragraph{Experimental setup} We repeat the following basic experiment five times with different random training/validation/test splits of the data. For each experimental repeat, we randomly select 80\% of the 922 patients to be in the training set and the remaining 20\% in the testing set. For Dynamic-DeepHit and DDRSA, within the training set, a random 20\% of the training points are held out as a validation set for hyperparameter tuning. The random splits are stratified as the preserve the fraction of data experiencing each event indicator value $\{0,1,\dots,k\}$. The hyperparameter grid we use is given in Appendix~\ref{sec:app-hyperparameters}. Note that the Fine and Gray model has no hyperparameters.

\subsection{Accuracy Benchmark in the Dynamic Problem Setup}
\label{sec:accuracy}

\paragraph{Survival analysis accuracy metric}
To evaluate the accuracy of different competing risks models, %
we compute c-indices per competing risk at different prediction times $t=6,12$ and different time horizons $\Delta=24,48,72$; these are all in units of hours. For example, with the prediction time $t=6$ and evaluation time $\Delta=24$, it means at hour 6 after ICU admission, we are comparing the model's predicted risk of different events occurring in the next 24 hours.
The results for the models are shown in Table~\ref{tab:c-index}. From this table, we see that in terms of c-indices per event, no model is uniformly the best across all prediction times $t$ and time durations $\Delta$, while DDRSA appears to have slightly lower accuracy scores compared to the other two models.

\begin{table}[t]
    \centering %
    \caption{Test set c-indices (average ± standard deviation across five experimental repeats) for three DCR models. The entries with bold values represent the highest average c-index for each $(\text{event},t,\Delta)$ combination.}\vspace{-1em}
    \adjustbox{scale=.8}{
    \begin{tabular}{cccccc} %
    \toprule
        \multirow{2.5}{*}{Model} & \multirow{2.5}{*}{Prediction time} & \multirow{2.5}{*}{Event} & \multicolumn{3}{c}{Evaluation time horizon} \\
        \cmidrule(lr){4-6}
        &  &  & $\Delta=24$ hrs & $\Delta=48$ hrs & $\Delta=72$ hrs \\ \midrule
        \multirow{6}{*}{Fine and Gray} & \multirow{3}{*}{$t=6$} & awakening & {\bftab 0.853 $\pm$ 0.017} & {\bftab 0.874 $\pm$ 0.012} & {\bftab 0.875 $\pm$ 0.012}\\ 
        & & death & 0.633 $\pm$ 0.171 & 0.673 $\pm$ 0.081 & 0.684 $\pm$ 0.061\\ 
        & & withdrawal & 0.691 $\pm$ 0.063 & {\bftab 0.634 $\pm$ 0.050} & {\bftab 0.652 $\pm$ 0.039}\\ \cmidrule(lr){2-6}
        & \multirow{3}{*}{$t=12$} & awakening & 0.831 $\pm$ 0.032 & 0.851 $\pm$ 0.023 & 0.854 $\pm$ 0.023\\ 
        & & death & {\bftab 0.751 $\pm$ 0.110} & {\bftab 0.709 $\pm$ 0.040} & 0.713 $\pm$ 0.044\\ 
        & & withdrawal & 0.709 $\pm$ 0.082 & {\bftab 0.675 $\pm$ 0.035} & {\bftab 0.681 $\pm$ 0.024}\\
        \midrule
        \multirow{6}{*}{Dynamic-DeepHit} & \multirow{3}{*}{$t=6$} & awakening & 0.851$\pm$0.018 & 0.867$\pm$0.012 & 0.864±0.017 \\
        & &  death & {\bftab 0.702±0.080} & {\bftab 0.684±0.096} & {\bftab 0.697±0.071}\\ 
        & & withdrawal & {\bftab 0.742±0.103} & 0.612±0.062 & 0.621±0.031\\ \cmidrule(lr){2-6}
        & \multirow{3}{*}{$t=12$} & awakening & {\bftab 0.847±0.038} & {\bftab 0.859±0.020} & {\bftab 0.858±0.024}\\ 
        & & death & 0.701±0.078 & 0.699±0.042 & {\bftab 0.722±0.044}\\ 
        & & withdrawal & {\bftab 0.739±0.076} & 0.640±0.028 & 0.666±0.017\\ 
        \midrule
        \multirow{6}{*}{DDRSA} & \multirow{3}{*}{$t=6$} & awakening & 0.821±0.032 & 0.845±0.028 & 0.836±0.030\\ 
        & & death & 0.599±0.072 & 0.615±0.059 & 0.633±0.075\\ 
        & & withdrawal & 0.677±0.151 & 0.626±0.121 & 0.637±0.100\\ \cmidrule(lr){2-6}
        & \multirow{3}{*}{$t=12$} & awakening & 0.825±0.021 & 0.818±0.025 & 0.798±0.027\\ 
        & & death & 0.681±0.095 & 0.651±0.086 & 0.651±0.074\\ 
        & & withdrawal & 0.663±0.126 & 0.652±0.092 & 0.670±0.068\\
        \bottomrule
    \end{tabular}}
    \label{tab:c-index} %
\end{table}

\vspace{-.2em}
\paragraph{Binary classification accuracy metric}
By using the binary classifier derived from a DCR model as described in Section~\ref{sec:classifier}, we can use binary classification evaluation metrics such as the area under the ROC curve (AUROC). After restricting the test set cohort to patients that we either observe awakening or death %
(not from withdrawal from life-sustaining therapies), we can compute the AUROC scores shown in Table~\ref{tab:auroc}. %
Here, note that the Fine and Gray model achieves mean AUROC scores that are higher than those of Dynamic-DeepHit across all values of $t$ and $\Delta$ evaluated. Accounting for the standard deviations of the AUROC scores, the two models do have AUROCs that are quite close. Again, DDRSA seems to perform worse than the other two models in terms of AUROCs.

\begin{figure}[t]
  \centering %
  \subfigure[$t=6, \Delta=24$]{\includegraphics[width=0.45\linewidth]{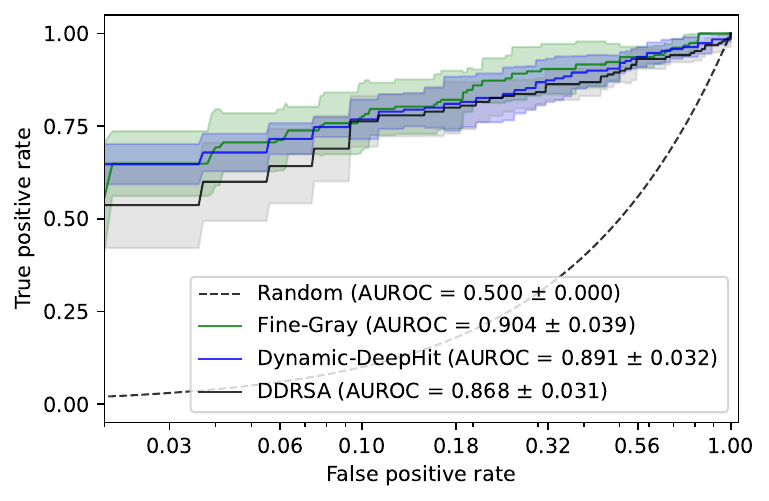}} \label{fig:roc-6-24}
  ~
  \subfigure[$t=12, \Delta=24$]{\includegraphics[width=0.45\linewidth]{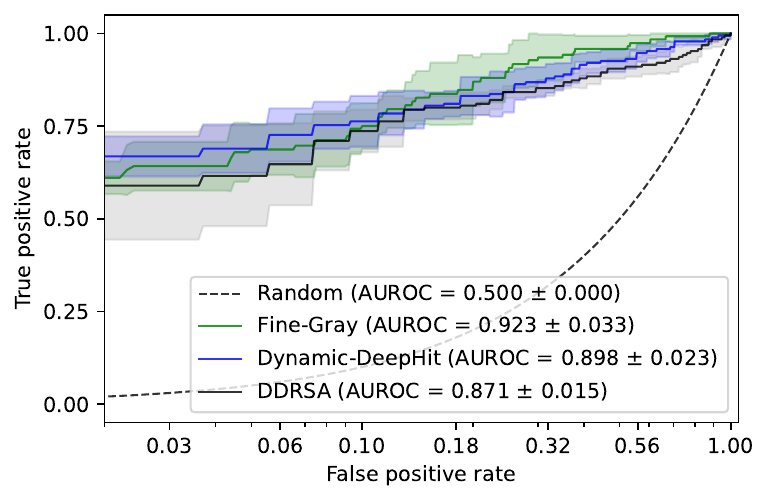}} \label{fig:roc-12-24}
  \vspace{-.75em}
  \caption{Test set ROC curve (average curve $\pm$ standard deviation intervals across five experimental repeats). The x-axis is on a log scale to emphasize the low FPR~regime.}
  \label{fig:roc}\vspace{-2em}
\end{figure}

We can also plot the ROC curve at different $t$ and $\Delta$ values. Specifically, we show the ROC at $t=6$ and $t=12$ with the estimated ratio of the probability of awakening and probability of death (not from withdrawal) within the next $\Delta=24$ hours in Figure~\ref{fig:roc}, where we plot the x-axis on a log scale to focus on the low FPR regime. For example, we can see that at $t=12, \Delta=24$, Dynamic-DeepHit reaches an average AUROC of 0.898 and an average TPR of 0.668 with a small FPR of 0.020. The ROC curves for a few other $t$ and $\Delta$ values can be found in Appendix~\ref{app-roc}.

\begin{table}[t]
    \centering %
    \caption{Test set AUROC (average ± standard deviation across five experimental repeats) for three DCR models. The entries with bold values represent the highest average AUROC for each $(t,\Delta)$ combination.}\vspace{-.9em}
    \adjustbox{scale=.8}{
    \begin{tabular}{ccccc} %
    \toprule
        \multirow{2.5}{*}{Model} & \multirow{2.5}{*}{Prediction time} & \multicolumn{3}{c}{Evaluation time horizon} \\
        \cmidrule(lr){3-5}
        &  &  $\Delta=24$ hrs & $\Delta=48$ hrs & $\Delta=72$ hrs \\ 
        \midrule
        \multirow{2}{*}{Fine and Gray} & $t=6$ & {\bftab 0.904 ± 0.039} & {\bftab 0.903 ± 0.039}  & {\bftab 0.903 ± 0.039} \\ 
        & $t=12$ & {\bftab 0.923 ± 0.033} & {\bftab 0.921 ± 0.034} & {\bftab 0.920 ± 0.034} \\
        \midrule
        \multirow{2}{*}{Dynamic-DeepHit} & $t=6$ &  0.891 ± 0.032 &  0.883 ± 0.030 &   0.885 ± 0.029\\ 
        & $t=12$ &  0.898 ± 0.023 &  0.891 ± 0.029 &  0.899 ± 0.025\\ 
        \midrule
        \multirow{2}{*}{DDRSA} & $t=6$ & 0.868 ± 0.031 & 0.863 ± 0.028 & 0.849 ± 0.031 \\ 
        ~ & $t=12$ & 0.871 ± 0.015 & 0.853 ± 0.019 & 0.822 ± 0.020 \\
        \bottomrule
    \end{tabular}}
    \label{tab:auroc}\vspace{-.5em}
\end{table}

\subsection{Patient-Specific Visualizations}
\label{sec:multiple-visualizations}
After we train a DCR model, we can easily derive the estimated CIF for different events at different prediction times (e.g., $t=6,12$), and the estimated conditional probability of awakening using equation~\eqref{eq:prob-awake}. We focus on using the trained Dynamic-DeepHit to derive all the visualizations in this part as an illustrative example (the same sorts of visualizations could be made for the Fine and Gray model and DDRSA). In Figure~\ref{fig:cif-eeg}, we display two summary EEG signals for four example patients, the estimated CIFs, and the heat map visualization we described in Section~\ref{sec:visualization}; each row/panel in the figure corresponds to one patient. %
Each entry in the heatmap is the estimated conditional probability of awakening occurring at different times ($t = 1,2,\dots,12$) for the different durations ($\Delta = 24, 48, 72$). For the four patients shown:
\begin{itemize}[itemsep=0pt,parsep=0pt,topsep=2pt]
    \item (Figure~\ref{fig:cif-eeg}$(a)$) Patient 1 died at hour 17 (not from withdrawal of life-sustaining therapies). We observe a drastic change in the patient's EEG signals before and after $t=6$, which is reflected in the two sets of estimated CIFs at $t=6$ and $t=12$ with a higher estimated CIF of the ``awakening'' event given the first 6 hours of EEG data but a lower ``awakening'' CIF curve after the first 12 hours of EEG.
    \item (Figure~\ref{fig:cif-eeg}$(b)$) Patient 2 is still in a coma at hour 118. From the EEG signals of this patient, we can see that the probability of awakening is increasing gradually in the heat map (fourth column plot) as we go from $t=1$ to $t=12$.
    \item (Figure~\ref{fig:cif-eeg}$(c)$) Patient 3 awakened at hour 54, and we do observe good EEG signals (we briefly describe some common patterns considered ``good'' or ``bad'' in the caption of Figure~\ref{fig:cif-eeg}). In the heat map visualization, the predicted probability of awakening is high at all times.
    \item (Figure~\ref{fig:cif-eeg}$(d)$) Patient 4 died (not from withdrawal of life-sustaining therapies) at hour 71, and we did observe very poor EEG signals over the entire 12-hour period. In the heat map, the predicted probability of awakening is low at all entries.
\end{itemize}
Note that the estimated CIF for withdrawal from life-sustaining therapies could be viewed as the model's prediction of how likely a physician (at least according to historical data) would make a decision to withdraw said therapies. %

\subsection{Ablation Study}
\label{sec:ablation}

We conduct an ablation study to show why including the three competing risks in how we framed the problem is better than had we used fewer competing risks. 
In particular, we repeat the same experiments as above but with only two competing risks (awakening and dying not by withdrawal from life-sustaining therapies), and a single event (awakening). For the purpose of this ablation study, we only focus on the Fine and Gray model and Dynamic-DeepHit as they achieved noticeably higher accuracy than DDRSA in our earlier experiments.

In the case when we only consider two competing risks, a patient with the outcome label of withdrawal from life-sustaining therapies would be viewed as being censored (along with those who stayed in a coma). Similarly, when we only model a single competing risk, patients with all other outcomes are viewed as being censored. The resulting c-indices are shown for the two-competing-risk case in Table~\ref{tab:c-index-2-event} and for the one competing risk case in Table~\ref{tab:c-index-1-event}.

\begin{table}[t]
    \centering %
    \caption{Test set c-indices (average $\pm$ standard deviation across five experimental repeats) for the Fine and Gray model and Dynamic-DeepHit with two competing events. The entries with bold values represent the highest average c-index for each \mbox{(event, $t, \Delta$)} combination.}\vspace{-.75em}
    \adjustbox{scale=.8}{
    \begin{tabular}{cccccc} %
    \toprule
        \multirow{2.5}{*}{Model} & \multirow{2.5}{*}{Prediction time} & \multirow{2.5}{*}{Event} & \multicolumn{3}{c}{Evaluation time horizon} \\
        \cmidrule(lr){4-6}
        & &  &  $\Delta=24$ hrs & $\Delta=48$ hrs & $\Delta=72$ hrs \\ \midrule
        \multirow{4}{*}{Fine and Gray} & \multirow{2}{*}{$t=6$} & awakening & 0.835 $\pm$ 0.030 & {\bftab 0.870 $\pm$ 0.012} & {\bftab 0.867 $\pm$ 0.008} \\ 
        & & death & {\bftab 0.723 $\pm$ 0.040} & {\bftab 0.697 $\pm$ 0.075} & {\bftab 0.707 $\pm$ 0.061} \\ 
        \cmidrule(lr){2-6}
        & \multirow{2}{*}{$t=12$} & awakening & 0.822 $\pm$ 0.020 & 0.855 $\pm$ 0.014 & 0.851 $\pm$ 0.014 \\ 
        & ~ & death & {\bftab 0.664 $\pm$ 0.272} & {\bftab 0.678 $\pm$ 0.146} & {\bftab 0.706 $\pm$ 0.101} \\
        \midrule
        \multirow{4}{*}{Dynamic-DeepHit} & \multirow{2}{*}{$t=6$} & awakening & {\bftab 0.852 $\pm$ 0.019} & 0.858 $\pm$ 0.011 & 0.855 $\pm$ 0.007 \\ 
        & & death & 0.638 $\pm$ 0.074 & 0.622 $\pm$ 0.073 & 0.642 $\pm$ 0.050 \\ 
        \cmidrule(lr){2-6}
        & \multirow{2}{*}{$t=12$} & awakening & {\bftab 0.842 $\pm$ 0.017} & {\bftab 0.860 $\pm$ 0.007} & {\bftab 0.857 $\pm$ 0.007} \\ 
        & ~ & death & 0.599 $\pm$ 0.092 & 0.639 $\pm$ 0.095 & 0.649 $\pm$ 0.078 \\
        \bottomrule
    \end{tabular}}
    \label{tab:c-index-2-event}
    \vspace{-1em}
\end{table}

\begin{table}[t]
    \centering %
    \caption{Test set c-indices (average $\pm$ standard deviation across five experimental repeats) for the Fine and Gray model and Dynamic-DeepHit with a single event. The entries with bold values represent the highest average c-index for each \mbox{(event, $t,\Delta$)} combination.}\vspace{-1em}
    \adjustbox{scale=.8}{
    \begin{tabular}{cccccc} %
    \toprule
        \multirow{2.5}{*}{Model} & \multirow{2.5}{*}{Prediction time} & \multirow{2.5}{*}{Event} & \multicolumn{3}{c}{Evaluation time horizon} \\
        \cmidrule(lr){4-6}
        & &  &  $\Delta=24$ hrs & $\Delta=48$ hrs & $\Delta=72$ hrs \\ \midrule
        \multirow{2}{*}{Fine and Gray} & $t=6$ & awakening & {\bftab 0.864 $\pm$ 0.021} & {\bftab 0.883 $\pm$ 0.013} & {\bftab 0.877 $\pm$ 0.010} \\ 
        \cmidrule(lr){2-6}
        & $t=12$ & awakening & 0.843 $\pm$ 0.023 & {\bftab 0.866 $\pm$ 0.018} & {\bftab 0.861 $\pm$ 0.015} \\ 
        \midrule
        \multirow{2}{*}{Dynamic-DeepHit} & $t=6$ & awakening & 0.850 $\pm$ 0.031 & 0.861 $\pm$ 0.019 & 0.855 $\pm$ 0.021 \\ 
        \cmidrule(lr){2-6}
        & $t=12$ & awakening & {\bftab 0.845 $\pm$ 0.021} & 0.855 $\pm$ 0.014 & 0.855 $\pm$ 0.018 \\ 
        \bottomrule
    \end{tabular}}
    \label{tab:c-index-1-event}\vspace{-.5em}
\end{table}

By comparing Tables~\ref{tab:c-index} and~\ref{tab:c-index-2-event}, we see that c-indices for death (not by withdrawal of life-sustaining therapies) stay at a similar level for the Fine and Gray model considering the standard deviation interval when we model only two competing risks, while those of Dynamic-DeepHit are clearly higher when we model all three competing risks.
While we do not observe a gain in the c-indices for the event of awakening when we move from the single risk setting to those of two or three competing risks, by incorporating more events, we are able to capture more information without decreasing the model's accuracy. For example, if we only trained the single competing risk model and if the patient is not likely to awaken from the coma, the model would not help us distinguish between any of the other possible outcomes of the patient.

\begin{figure}[t]
  \centering %
  \subfigure[$t=6, \Delta=24$]{\includegraphics[width=0.45\linewidth]{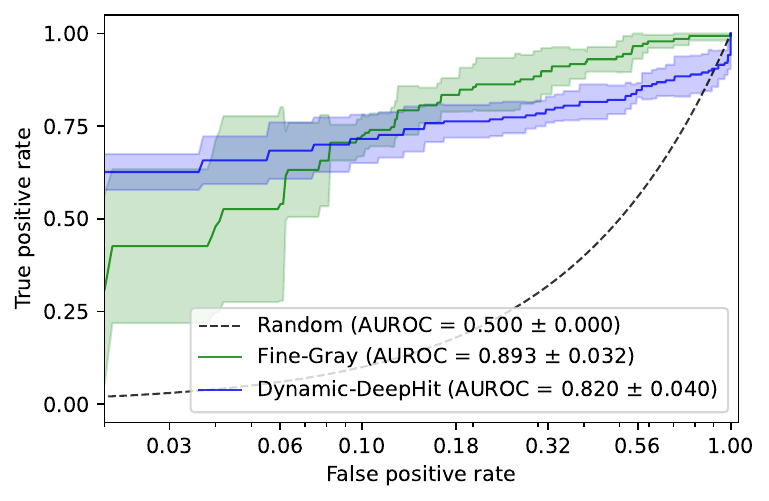}} \label{fig:roc-6-24-cr2}
  ~
  \subfigure[$t=12, \Delta=24$]{\includegraphics[width=0.45\linewidth]{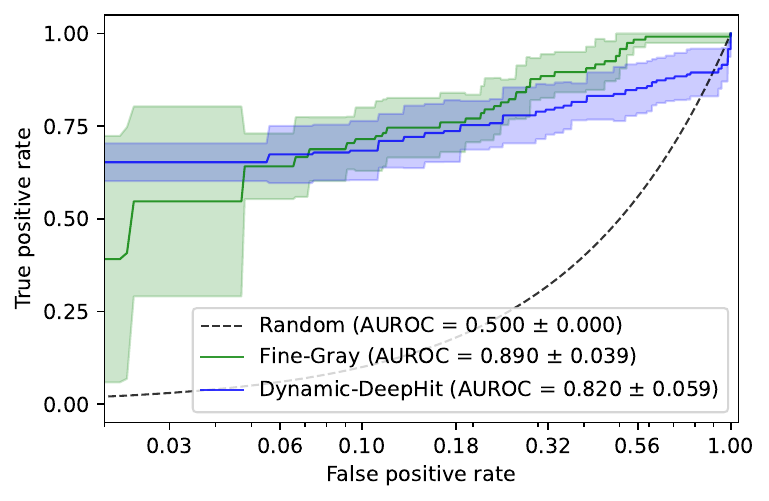}} \label{fig:roc-12-24-cr2}
  \vspace{-.75em}
  \caption{Test set ROC curve in the case of modeling two competing events (average curve $\pm$ standard deviation intervals across five experimental repeats). The x-axis is on a log scale to emphasize the low FPR~regime.}
  \label{fig:roc-cr2} %
\end{figure}

We also derive the binary classifier under two competing risks. The resulting AUROC scores are shown in Table~\ref{tab:auroc-cr2} and ROC plots at $t=6,12, \Delta=24$ in Figure~\ref{fig:roc-cr2}. By comparing Tables~\ref{tab:auroc} and \ref{tab:auroc-cr2}, average AUROCs drop for both the Fine and Gray model and Dynamic-DeepHit when we change from three events to two events, showing that there is a benefit to including the event of withdrawal from life-sustaining therapies. By comparing Figures~\ref{fig:roc} and \ref{fig:roc-cr2}, focusing on the low FPR regime, we observe the TPRs for Dynamic-DeepHit stay at a similar level under the two event setting while those for the Fine and Gray model become very unstable, again, suggesting the benefit of including the event of withdrawal. The ROC curves under the two event setting for a few other $t$ and $\Delta$ values can be found in Appendix~\ref{app-ablation}.

\begin{table}[t]
    \centering %
    \caption{Test set AUROC (average ± standard deviation across five experimental repeats) for the Fine and Gray model and Dynamic-DeepHit with two events. The entries with bold values represent the highest average AUROC for each $(t,\Delta)$ combination.}\vspace{-1em}
    \adjustbox{scale=.8}{
    \begin{tabular}{ccccc} %
    \toprule
        \multirow{2.5}{*}{Model} & \multirow{2.5}{*}{Prediction time} & \multicolumn{3}{c}{Evaluation time horizon} \\
        \cmidrule(lr){3-5}
        &  &  $\Delta=24$ hrs & $\Delta=48$ hrs & $\Delta=72$ hrs \\ 
        \midrule
        \multirow{2}{*}{Fine and Gray} & $t=6$ & {\bftab 0.893 ± 0.032} & {\bftab 0.892 ± 0.033} &  {\bftab 0.891 ± 0.033} \\ 
        & $t=12$ & {\bftab 0.890 ± 0.039} & {\bftab 0.887 ± 0.038} & {\bftab 0.886 ± 0.038} \\
        \midrule
        \multirow{2}{*}{Dynamic-DeepHit} & $t=6$ & 0.820 ± 0.040  & 0.827 ± 0.034 &  0.832 ± 0.035 \\ 
        & $t=12$ &  0.820 ± 0.059 &  0.824 ± 0.053 &  0.835 ± 0.057\\ 
        \bottomrule
    \end{tabular}}
    \label{tab:auroc-cr2} %
\end{table}

\section{Discussion} \label{sec:discussion}

Our paper proposes a dynamic formulation of the neurological prognostication problem for post-cardiac-arrest coma patients. The modeling solution we propose uses any existing DCR model and, using specific structure in clinical outcomes of our clinical application, derives a classifier from the DCR model that aims to be helpful to clinicians for decision support. We believe that our dynamic formulation better models the specific clinical problem we focus on compared to what has been proposed in existing literature.

We now discuss an alternative to our conditioning strategy in Section~\ref{sec:classifier} that we believe is worth investigating in future work, and we also point out various other limitations.

\paragraph{An alternative solution for deriving a binary classifier}
In how we derived our binary classifier in Section~\ref{sec:classifier}, we conditioned on the event that \emph{the test patient never gets withdrawn from life-sustaining therapies} (the ``non-withdrawal event''). Our classifier uses the probability of awakening and, separately, of dying both conditioned on the non-withdrawal event. However, we have found that these conditional probabilities are not entirely straightforward to explain to practitioners. Part of the challenge is that it is unclear how conditioning on the non-withdrawal event should impact the probabilities estimated. Concretely, consider the probability of awakening. Prior to conditioning on the non-withdrawal event vs after conditioning on the event, it is unclear whether the probability of awakening should increase or decrease.

Fundamentally, two technical hurdles make reasoning about the non-withdrawal event difficult. First, we do not know what would have happened to patients withdrawn from life-sustaining therapies had they been kept on these therapies instead. Second, we do not know how ``good'' past decisions on withdrawing life-sustaining therapies are. As an extreme example, suppose that 100\% of patients who die from non-withdrawal causes are perfectly identified by physicians in advance that they would have no chance of awakening so they are pulled off life-sustaining therapies, and 100\% of patients who would awaken are also perfectly identified by physicians so that they are kept on life-sustaining therapies. In this case, if we condition on the non-withdrawal event, then we would always just get that the conditional probability of awakening is 100\%, so our binary classifier in Section~\ref{sec:classifier} would not be useful. However, an alternative that would still be useful is to compute the probabilities of awakening and of dying (not of withdrawal from life-sustaining therapies) \emph{without} conditioning on the non-withdrawal event. These probabilities would still be conditional probabilities since we would condition on the test patient's time series. However, we no longer condition on the non-withdrawal event. We now sketch an approach for computing these probabilities, which in turn leads to a binary classifier different from the one we derived in Section~\ref{sec:classifier}.

\begin{figure}[t]
  \centering
  \subfigure[Example Patient 1]%
  {\includegraphics[width=0.99\linewidth]{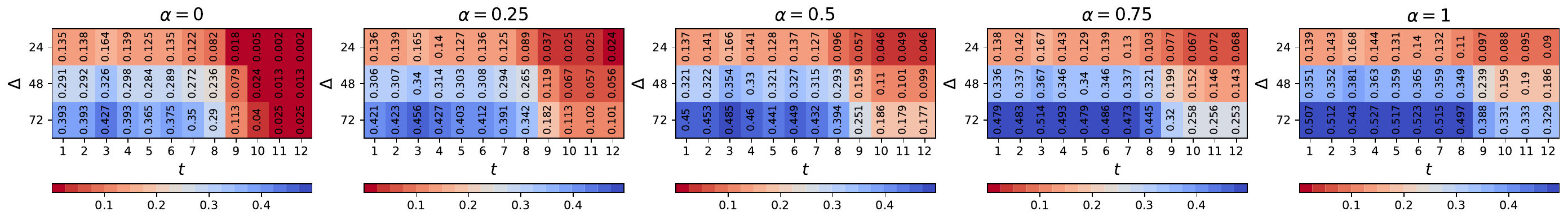}\label{fig:p1-visual2}}
  \vspace{-.75em}
  \caption{For the same Example Patient 1 in panel $(a)$ of Figure~\ref{fig:cif-eeg}, we show the probability of awakening $\widetilde{P}_{\text{awaken}}(\Delta\mid z, t)$ (from equation~\eqref{eq:prob-awaken-alternative}) for different values of $\alpha$.} %
  \label{fig:prob-awake}
  \vspace{-2em}
\end{figure}

To begin with, let's consider the probability of awakening within time duration $\Delta$ for a patient with features observed until time $t$. We write this probability as
\begin{align}
&\widetilde{P}_{\text{awaken}}(\Delta \mid z, t) \nonumber\\
&= \mathbb{P}(\text{earliest competing event that happens excluding withdrawal from life-sustaining}
\nonumber\\
&\phantom{= \mathbb{P}(~\!}\text{therapies is awakening}, Y\le t + \Delta| Z^{(\le t)} = z^{(\le t)}, Y>t)
\nonumber\\
&= \mathbb{P}(\text{earliest competing event that happens is awakening}, Y\le t + \Delta| Z^{(\le t)} = z^{(\le t)}, Y>t)
\nonumber\\
&\phantom{=}+\mathbb{P}(\text{earliest competing event that happens is withdrawal from life-sustaining} \nonumber\\
&\phantom{= + \mathbb{P}(~\!}\text{therapies followed by awakening}, Y\le t + \Delta| Z^{(\le t)} = z^{(\le t)}, Y>t). \nonumber
\label{eq:prob-awaken-broken}
\end{align}
On the right-hand side above, the first term is just the CIF for awakening (i.e., $F_1(\Delta\mid z, t)$ using equation~\eqref{eq:CIF} and the same notation as in Section~\ref{sec:classifier}). As for the second term, we now make a major assumption that
\begin{align}
&\mathbb{P}(\text{earliest competing event that happens is withdrawal from life-sustaining} \nonumber\\
&\phantom{\mathbb{P}(}\text{therapies followed by awakening}, Y\le t + \Delta| Z^{(\le t)} = z^{(\le t)}, Y>t) \nonumber \\
&\quad = F_3(\Delta\mid z, t) \times \alpha, \nonumber
\end{align}
where $\alpha\in[0,1]$ is a constant that does not depend on anything else, and as a reminder $F_3$ is the CIF for withdrawal from life-sustaining therapies. The above assumption says that among patients whose earliest competing event is being withdrawn from life-sustaining therapies, each of them has probability $\alpha$ (independent of everything else) of having their second earliest competing event be awakening (so that had withdrawal from life-sustaining therapies not been an option, they would have awaken). In practice, we have no way of estimating $\alpha$ but we can try different values for it. Putting together the pieces, we have
\begin{equation}
\widetilde{P}_{\text{awaken}}(\Delta \mid z, t)
= F_1(\Delta\mid z, t) + F_3(\Delta\mid z, t)\times\alpha.
\label{eq:prob-awaken-alternative}
\end{equation}
In Figure~\ref{fig:prob-awake}, we provide the heatmap visualization of the probability of awakening with $\alpha\in\{0,0.25,0.5,0.75,1\}$ for the same Example Patient 1 as in Figure~\ref{fig:cif-eeg}. We can see when $\alpha$ increases from zero to one, the estimation becomes more ``optimistic'' in terms of the probability of awakening being higher across different times $t$ and durations $\Delta$.

The same logic can be used to derive the probability of dying (of causes aside from withdrawal from life-sustaining therapies) within duration $\Delta$; we can denote this probability as $\widetilde{P}_{\text{death (not withdrawal)}}(\Delta\mid z, t)$. Then we can derive a binary classifier in the same manner as how we derived the one in Section~\ref{sec:classifier}: e.g., we threshold on the ratio $\frac{\widetilde{P}_{\text{death (not withdrawal)}}(\Delta \mid z, t)}{\widetilde{P}_{\text{awaken}}(\Delta \mid z, t)}$ to decide on whether to predict between awakening or dying.

At present, we do not yet fully understand when our proposed solution in Section~\ref{sec:classifier} should be used in practice vs the one stated in this discussion section which makes the major assumption involving the unknown probability $\alpha$ of patients withdrawn from life-sustaining therapies who would have awaken. Better understanding the pros and cons of these approaches would be interesting. Relaxing the factorization assumption involving the probability $\alpha$ would also be an interesting future research direction.

\paragraph{Other limitations}
At a high-level, our paper has proposed a framework in which to think about the neurological prognostication problem that works with any DCR model. However, at this point, we have not proposed a new DCR model and, as far as we know, there have simply not been many DCR models developed (where the model truly is using variable-length time series rather than how we set up the Fine and Gray baseline which actually only uses fixed-length feature vector inputs). Naturally, a future research direction would be the development of DCR models that are even more accurate than the ones we have tested, especially since for the accuracy metrics we use, state-of-the-art neural network DCR models do not appear to be significantly better than the classical Fine and Gray model that only uses the final time step's features (including static features).

In terms of evaluating the binary classifier, we simply use AUROC scores at different $t, \Delta$ with patients who we either observed awakening or death (not from the withdrawal of life-sustaining therapy), ignoring those who have been withdrawn or censored due to lost of follow-up, which could be biased. There are numerous work in the area of ROC analysis with the occurrence of censoring under time-dependent setting (e.g., \citealt{blanche2013estimating, kamarudin2017time, li2018simple}), each with its own limitations. While it is not the focus of our current work, we encourage future work to probe how to better evaluate the binary classifier.

Another limitation of our framework is that we only consider the first ``hitting time'' of competing risks, i.e. the earliest event that happened. In reality, some patients may still experience unfavorable outcomes after awakening, and it could be important to try to predict when this happens. One simple approach would be to split up the ``awaken'' event into different types of ``awaken'' events, although perhaps a better modeling framework would be to consider multiple critical events that could happen, one after another, rather than constraining the setting to only be on first hitting times. %

As for the specific clinical problem we have focused on, there were a number of limitations related to the dataset we curated.
First, the EEG data we have is limited to the first 12 hours after ICU admission. If we were to have more complete EEG data, we would potentially be able to provide more accurate prognoses. Second, for ease of computation, we downsample the EEG time series data. From some preliminary analyses, we found that this did not impact the resulting prediction accuracy much. However, more thorough experiments are needed to better understand the impact of our current downsampling procedure on information loss. We remark that our patient heat map visualization can help us also find when our DCR-derived classifier is actually wrong but it is very confident in its wrong answer (we provide some examples of this in Appendix~\ref{app-visual}). In some such cases, from physician feedback, we have learned that the prognostication task even for physicians could be difficult given only EEG data and the static features we use. In particular, %
by collecting and using additional patient features (e.g., vitals, whether the patient experienced another cardiac arrest), more accurate predictions should be possible. %

Ultimately, although we believe that our paper takes a step toward more realistically framing the problem of neurological prognostication of post-cardiac-arrest coma patients compared to existing literature, our work has not yet led to a ``deployment-ready'' solution. Moreover, at this point, it is unclear to what extent the patient-specific heat map visualization we proposed %
actually helps clinical decision support. By continuing to account for physician feedback, we hope that we can produce a decision support system that is practically useful. User studies with clinicians would be required to assess the impact of such a system on clinical decision making.

\section*{Acknowledgments} \label{sec:ack}
This work was supported by NSF CAREER award \#2047981. The authors thank the anonymous reviewers for helpful feedback.

\bibliography{bib}

\newpage
\appendix
\counterwithin{table}{section}
\counterwithin{figure}{section}

\section{Data} \label{sec：app-data}

Some summary statistics for the dataset corresponding to the cohort described in Section~\ref{sec:cohort} are provided in Table~\ref{tab:summary-stats}.

\paragraph{EEG Features}
After post-cardiac-arrest comatose patients are admitted to the intensive care unit, electroencephalography (EEG) is used to measure brain activity. For the data we use, EEG signals are typically recorded at 256Hz from 22 electrodes adhering to standard positions according to the 10-20 International System of electrode placement. 
They are then processed via the medical software Persyst (\url{https://www.persyst.com/}) to generate 6,037 features at 1Hz, which is the ``raw'' format of the data we start with (note that the actual raw data originally recorded by the EEG electrodes are not available, only this processed version from Persyst). The 6,037 features from Persyst include time, artifact intensity, electrode signal quality, 
seizure probability, FFT (fast Fourier transform) spectrogram, aEEG, peak envelope (0-25 Hz), rhythmicity spectrogram, asymmetry EASI/REASI, relative asymmetry spectrogram, spikes, and suppression ratio. %
Out of the above features, for simplicity, we focus on two types of EEG features that have been previously proven to be informative for neurological prognostication (e.g., \citealt{oh2015continuous,elmer2016group}):

\textbf{Suppression ratio} is calculated as a 10-second summary and reflects the proportion of the amplitude of the EEG signals in that window that falls below some threshold compared to that which falls above the threshold. More suppressed EEG is a sign of more severe dysfunction or injury. Specifically, we use 2 features: the mean values of all the electrodes in the left and right hemispheres respectively.

\textbf{Amplitude-integrated EEG (aEEG)} is one way of describing the overall EEG amplitude. Very low amplitude EEGs are a sign of injury or dysfunction. Specifically, we use 10 features: the mean values of all the electrodes in the left and right hemispheres, each with 5 different statistical summaries (max, min, median, 25\% percentile, 75\% percentile).

In summary, we use the below 12 features in our study:

  \begin{itemize}[itemsep=0pt,parsep=0pt,topsep=2pt]
      \item Mean of suppression ratio of all the electrodes in the left hemisphere
      \item Mean suppression ratio of all the electrodes in the right hemisphere
      \item Mean of max aEEG values of all the electrodes in the left hemisphere
      \item Mean of min aEEG values of all the electrodes in the left hemisphere
      \item Mean of median aEEG values of all the electrodes in the left hemisphere
      \item Mean of 25\% percentile aEEG values of all the electrodes in the left hemisphere
      \item Mean of 75\% percentile aEEG values of all the electrodes in the left hemisphere
      \item Mean of max aEEG values of all the electrodes in the right  hemisphere
      \item Mean of min aEEG values of all the electrodes in the right  hemisphere
      \item Mean of median aEEG values of all the electrodes in the right  hemisphere
      \item Mean of 25\% percentile aEEG values of all the electrodes in the right  hemisphere
      \item Mean of 75\% percentile aEEG values of all the electrodes in the right hemisphere
      
  \end{itemize}

\begin{table}
    \centering %
    \caption{Summary statistics of patient characteristics and outcomes. %
            All features from ``Age (yr)'' and below in the table have values reported as means except for the different ``Presenting rhythm'' features, which are all in raw counts. In the column names, ``withdrawal'' is an abbreviation for withdrawal from life-sustaining therapies (due to poor perceived neurological prognosis).}
            \vspace{-.75em}
    \adjustbox{scale=.8}{
    \begin{tabular}{lccccc}
    \toprule
        ~ & \multirow{2}{*}{Total} & \multirow{2}{*}{Awakened} & Death (not from & \multirow{2}{*}{Withdrawal} & \multirow{2}{*}{Coma} \\ 
        ~ & & &  withdrawal) & & \\ \midrule
        Number of patients & 922 & 271 & 189 & 432 & 30 \\ 
        Percentage of patients & 100\% & 29.4\% & 20.5\% & 46.9\% & 3.3\% \\ \midrule
        Age (yr) & 57.3 & 55.8 & 54.6 & 59.4 & 57.8 \\ 
        Female sex & 0.37 & 0.35 & 0.44 & 0.36 & 0.40 \\
        Arrest out-of-hospital & 0.85 & 0.79 & 0.85 & 0.87 & 0.90 \\ 
        Presenting rhythm & ~ & ~ & ~ & ~ & ~ \\ 
        ~~~~Ventricular tachycardia/fibrillation & 284 & 140 & 38 & 98 & 8 \\ 
        ~~~~Pulseless electrical activity & 317 & 83 & 72 & 153 & 9 \\ 
        ~~~~Asystole & 268 & 40 & 63 & 153 & 12 \\ 
        ~~~~Unknown & 53 & 8 & 16 & 28 & 1 \\ 
        Arrest duration (min) & 20.4 & 13.3 & 26.4 & 22.4 & 18.1 \\ 
        Time to event (hr) & 113.0 & 78.9 & 90.6 & 111.5 & 583.8 \\ \bottomrule
    \end{tabular}}
    \label{tab:summary-stats}
\end{table}

\paragraph{Static Features}
A full list of the static features we use in the study and possible values they could take:
\begin{itemize}[itemsep=0pt,parsep=0pt,topsep=2pt]
    \item Demographic
        \subitem \texttt{age}: Age
        \subitem \texttt{female}: Female gender
    \item Heart arrest
        \subitem \texttt{oohca}: Arrest location. \textit{Out-of-hospital, or In-hospital}
        \subitem \texttt{edarrest}: Initial arrest occurred after ED arrival? \textit{Yes, or No}
        \subitem \texttt{rhythm}: Initial arrest rhythm. \textit{No loss of pulse, VT/VF, PEA, Asystole, or Unknown}
        \subitem \texttt{ca\_type}: Pittsburgh Cardiac Arrest Category, \textit{I, II, III, IV, or Unknown}
        \subitem \texttt{transfer}: Referral from outside facility? \textit{Yes, or No}
        \subitem \texttt{witnessed}: Witnessed arrest? \textit{No, Lay person witnessed, or EMS witnessed}
        \subitem \texttt{bystander\_cpr}: Bystander CPR? \textit{No, Lay person, or Professional}
        \subitem \texttt{shocks}: Number of AED and ALS shocks for the duration of the initial resuscitation
        \subitem \texttt{duration}: Estimated cumulative duration of CPR in minutes
    \item Initial coma status
        \subitem \texttt{four\_r\_0}: FOUR Score - Respiratory
        \subitem \texttt{four\_eye\_0}: FOUR Score - Eyes
        \subitem \texttt{four\_m\_0}: FOUR Score - Motor
        \subitem \texttt{pupils\_0}: Pupils status. \textit{Both reactive, One reactive, Both nonreactive or Unable to determine}
        \subitem \texttt{corneals\_0}: Corneals. \textit{Both present, One present, Neither present, or Unable to determine}
        \subitem \texttt{cough\_0}: Cough. \textit{Present, Absent, or Unable to determine}
        \subitem \texttt{gag\_0}: Gag. \textit{Present, Absent, or Unable to determine}
    \item Medical history: \textit{Yes, or No}
        \subitem \texttt{ccimi}: History of myocardial infarction
        \subitem \texttt{ccipvd}: History of peripheral vascular disease
        \subitem \texttt{ccidementia}: History of dementia
        \subitem \texttt{ccicva}: History of stroke or TIA
        \subitem \texttt{ccihemi}: History of hemiplegia
        \subitem \texttt{ccichf}: History of congestive heart failure
        \subitem \texttt{ccicvd}: History of cerebrovascular disease
        \subitem \texttt{ccicld}: History of chronic lung disease/COPD
        \subitem \texttt{ccictd}: History of connective tissue disease
        \subitem \texttt{ccipud}: History of peptic ulcer disease
        \subitem \texttt{cciaids}: History of AIDS
        \subitem \texttt{ccickd}: History of moderate to severe chronic kidney disease
        \subitem \texttt{ccielsd}: History of chronic liver disease
        \subitem \texttt{ccidm}: History of diabetes
        \subitem \texttt{ccica}: History of solid tumor
        \subitem \texttt{ccileukemia}: History of leukemia
        \subitem \texttt{ccilymphoma}: History of lymphoma
\end{itemize}

\section{Training the Subdistribution Hazard Model} \label{sec:app-fine-and-gray}
Since the subdistribution hazard model by \citet{fine1999proportional} does not take vary-length time series data as inputs, we instead only use the static features and the EEG features of the last hour available for each patient in the training set to estimate the parameters of the subdistribution hazard model. After the parameters are estimated, we then plug in the static features and the EEG features at $t=6$ to estimate the CIFs and calculate the c-indices corresponding to $t=6$ and $\Delta=24,48,72$. Similarly, we generate the c-indices corresponding to $t=12$ and $\Delta=24,48,72$, by using the same static features and EEG features at $t=12$. This process is repeated five times with different random splits of training and testing sets to get the average and standard deviation of c-indices as in Table~\ref{tab:c-index}. Note that there is no validation set used for hyperparameter tuning as there are no hyperparameters.

\section{Modifying DDRSA to Support Competing Risks} \label{sec:app-ddrsa-competing-risk}
To make DDRSA \citep{venkata2022intervene} support competing risks, we modify the model structure of the original DDRSA, largely inspired by Dynamic-DeepHit \citep{lee2019dynamic}. In the original DDRSA, the encoder RNN module takes in time-varying input features, and the decoder RNN module recurrently produces predicted hazards at different discretized time intervals. To make it support competing events, instead of having only one decoder RNN module, we have multiple decoder RNN modules to generate predictions for each of the competing events respectively. Instead of predicting hazards at different discretized time intervals, we make the model to predict the probability mass function variant of the CIF as in equation~\eqref{eq:CIF-pmf}, where a softmax layer is applied at the end to make sure the constraint in equation~\eqref{eq:sum-to-1-constraint} is satisfied. We also incorporate the attention mechanism as in Dynamic-DeepHit. The modified DDRSA can be trained in the same fashion as Dynamic-DeepHit.

\section{Hyperparameter Grid} \label{sec:app-hyperparameters}
Note that for Dynamic-DeepHit, the loss is of the form:
\[
\mathcal{L}_{\text{total}} = \mathcal{L}_1 + \alpha \cdot \mathcal{L}_2 + \beta \cdot \mathcal{L}_3
\]
where $L_1$ is the negative log likelihood loss term (this basically encourages first hitting times to be correctly predicted, accounting for censoring), $L_2$ is a ranking loss (especially as the standard survival analysis metric of concordance index is a ranking-based metric, having a ranking loss that correctly orders patients based on their predicted CIFs can be helpful), $L_3$ is the RNN prediction loss (recall that we ask the RNN to try to predict the next time step's feature vector), and lastly $\alpha > 0$ and $\beta > 0$ are hyperparameters. %

We use the following hyperparameter grid for both Dynamic-DeepHit and DDRSA:
\begin{itemize}[itemsep=0pt,parsep=0pt,topsep=2pt]
\item learning rate $\in \{10^{-4}, 5\times10^{-4}, 10^{-3}\}$
\item weights of different loss terms: $\alpha \in \{0.5, 1, 5\}, \beta \in \{0.05, 0.1, 0.5\}$
\item dropout rate $\in \{0.2, 0.4\}$
\end{itemize}
For all sets of hyperparameters, we train with the Adam optimizer and a batch size of 32. The maximum number of epochs we train the models is 100 while we stop training if the average concordance index (across all the events with $\Delta=24,48,72$) on the validation set does not improve for 10 epochs. The best hyperparameter set is chosen to be the one with the highest average concordance index (across all the events with $\Delta=24,48,72$) on the validation set.

\section{Additional ROC Curves}\label{app-roc}

\begin{figure}[t]
  \centering\vspace{.5em}
  \subfigure[$t=6, \Delta=48$]{\includegraphics[width=0.45\linewidth]{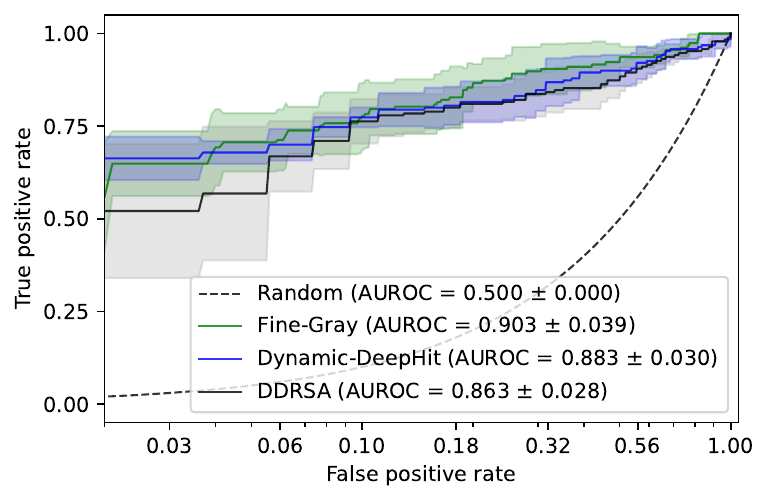}} \label{fig:roc-6-48}
  ~
  \subfigure[$t=12, \Delta=48$]{\includegraphics[width=0.45\linewidth]{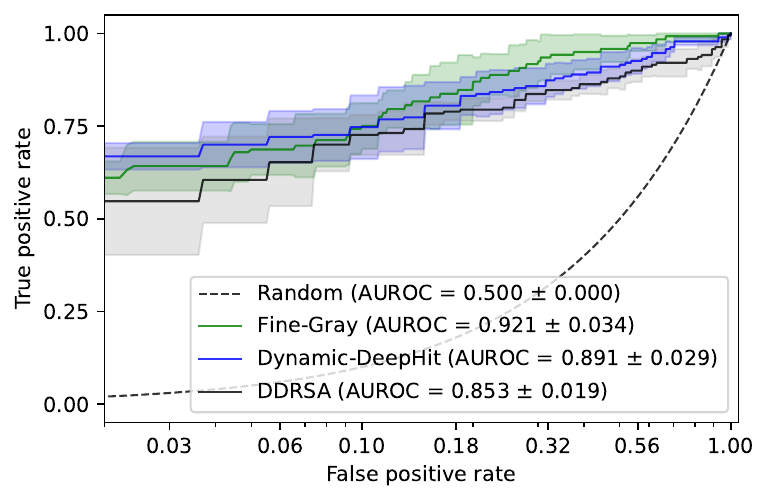}} \label{fig:roc-12-48}
  \vspace{-.5em}
  \vfill
  \centering\vspace{.5em}
  \subfigure[$t=6, \Delta=72$]{\includegraphics[width=0.45\linewidth]{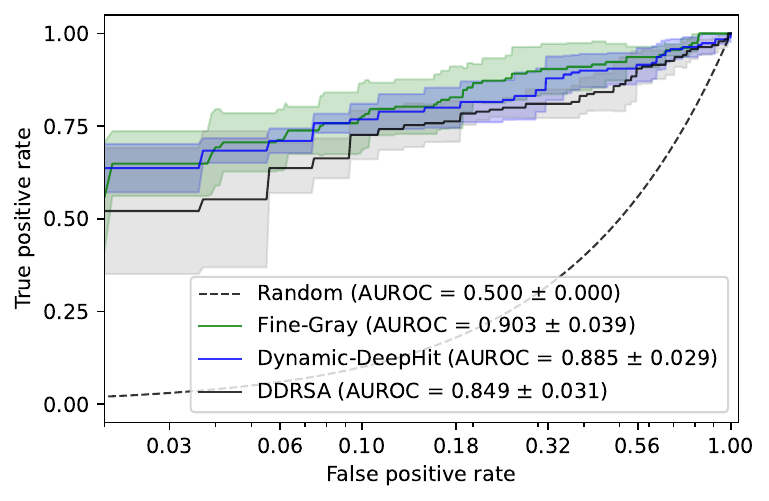}} \label{fig:roc-6-72}
  ~
  \subfigure[$t=12, \Delta=72$]{\includegraphics[width=0.45\linewidth]{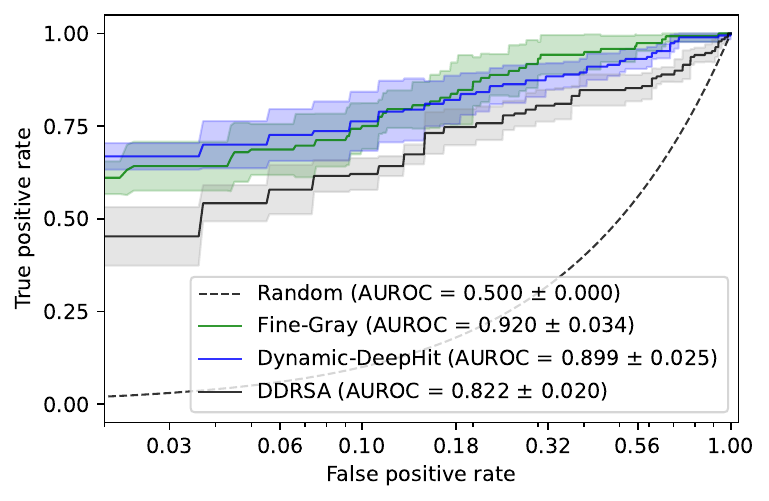}} \label{fig:roc-12-72}
  \vspace{-.5em}
  
  \caption{Test set ROC curve (average curve $\pm$ standard deviation intervals across five experimental repeats). The x-axis is on a log scale to emphasize the low FPR~regime.}
  \label{fig:roc-app}\vspace{-1.5em}
\end{figure}

We provide the ROC derived based on the estimated conditional probability of awakening for each patient at $t=6,12$ and $\Delta=48,72$, as shown in Figure~\ref{fig:roc-app}.

\section{Additional Patient-Specific Visualizations}\label{app-visual}
\begin{figure}[t]  %
  \centering
  \subfigure[Example Patient A: died (not from withdrawal of life-sustaining therapies) at hour 106 \label{fig:error-p1-visual}]{\includegraphics[width=0.99\linewidth]{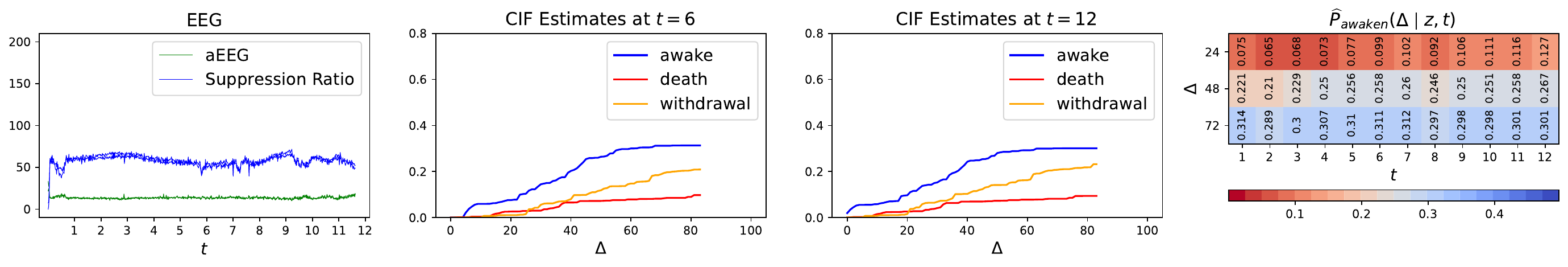}}
  \vfill
  \subfigure[Example Patient B: withdrawn from life-sustaining therapies at hour 101 \label{fig:error-p2-visual}]{\includegraphics[width=0.99\linewidth]{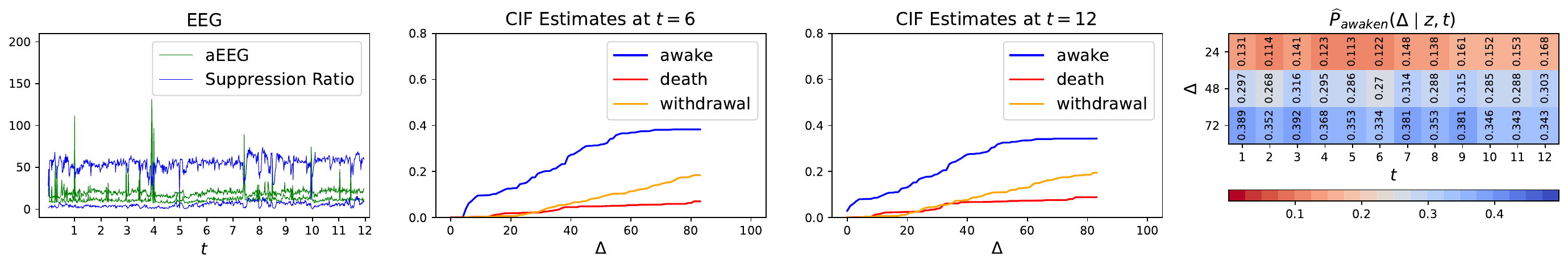}}%
  \vfill
  \subfigure[Example Patient C: died (not from withdrawal of life-sustaining therapies) at hour 14 \label{fig:error-p3-visual}]{\includegraphics[width=0.99\linewidth]{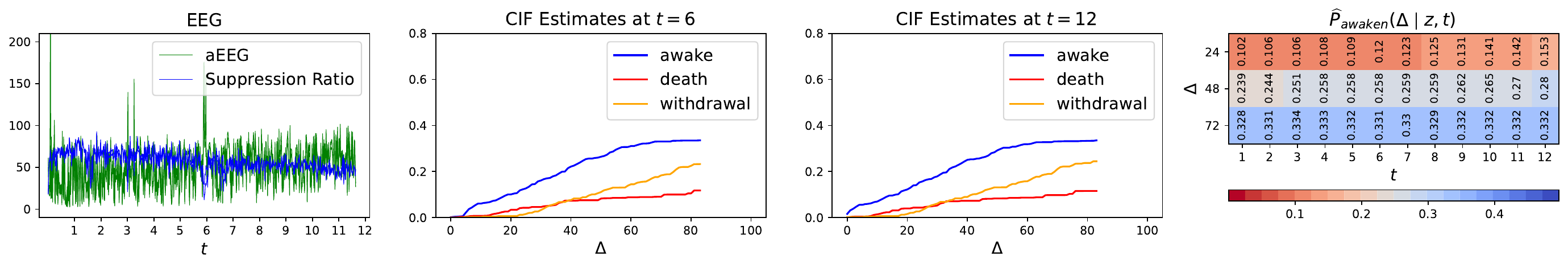}} %
  \vspace{-.75em}
  \caption{For each example patient (panels $(a)$-$(c)$), we show time series of two summary EEG features \emph{aEEG} and \emph{suppression ratio} (first column plot), estimated CIFs at hours $t=6$ (second column plot) and $t=12$ (third column plot), and our proposed heat map visualization (fourth column plot). Note that aEEG values for normal brain activity should be within a certain range (constantly being lower than 5 or higher than 25 is usually considered abnormal), and higher suppression ratio values are a sign of more severe dysfunction or injury.}
  \label{fig:error-cif-eeg}
\end{figure}

We show visualizations for three patients %
(Figure~\ref{fig:error-cif-eeg}) where the predictions are, in some sense, inconsistent with the EEG signals. %
In particular, for all three patients shown, their EEG signals would be perceived as corresponding to poor neurological activities, while the predictions of awakening are fairly high. While we do not understand what exactly leads to the overly high predicted probabilities of awakening for Patients A and C, we suspect the ``rareness'' of the EEG patterns is the reason for the inaccurate prediction for Patient B. In fact, the suppression ratio for one hemisphere and aEEG values are quite normal but the suppression ratio values for the other hemisphere are very high, indicating abnormal brain activities. This kind of inconsistency among different EEG features is very rare in the dataset that we curated.

\section{Additional ROC in Ablation Study}\label{app-ablation}
\begin{figure}[t]
  \centering\vspace{.5em}
  \subfigure[$t=6, \Delta=48$]{\includegraphics[width=0.45\linewidth]{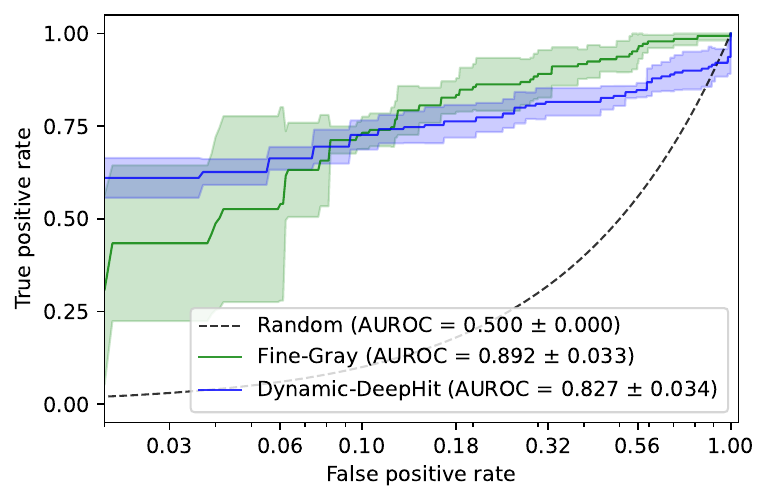}} \label{fig:roc-6-48-cr2}
  ~
  \subfigure[$t=12, \Delta=48$]{\includegraphics[width=0.45\linewidth]{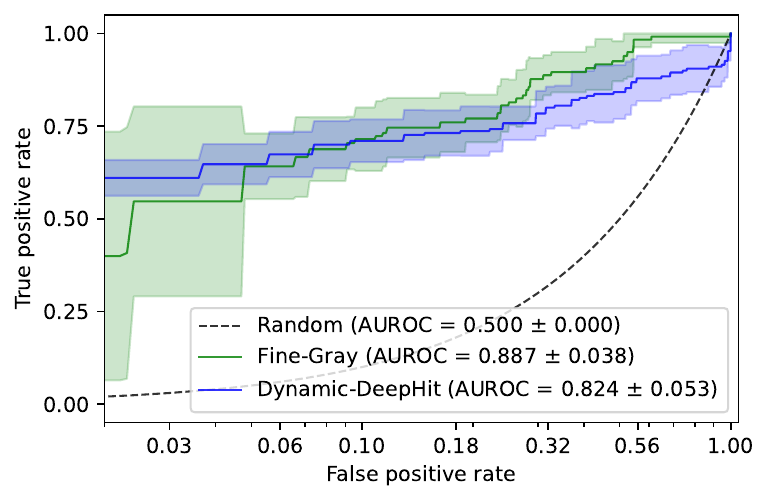}} \label{fig:roc-12-48-cr2}
  \vspace{-.5em}
  \vfill
  \centering\vspace{.5em}
  \subfigure[$t=6, \Delta=72$]{\includegraphics[width=0.45\linewidth]{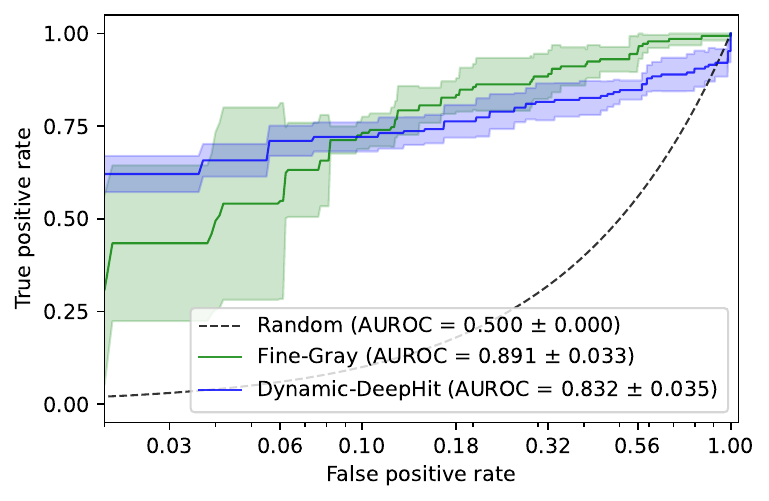}} \label{fig:roc-6-72-cr2}
  ~
  \subfigure[$t=12, \Delta=72$]{\includegraphics[width=0.45\linewidth]{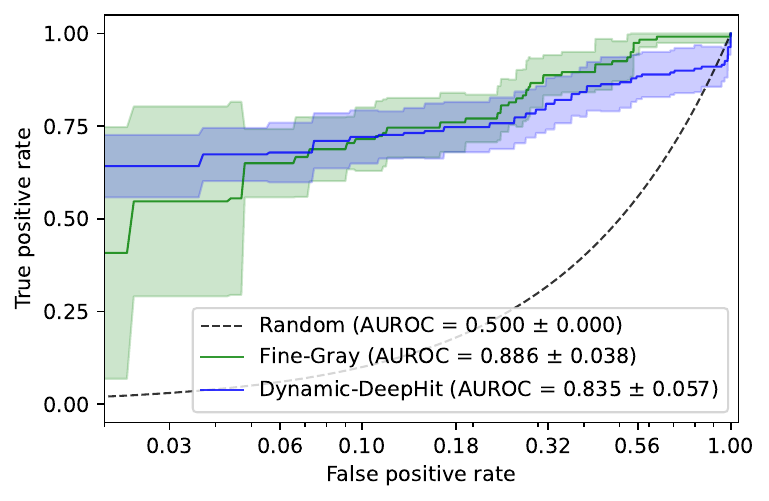}} \label{fig:roc-12-72-cr2}
  \vspace{-.5em}
  
  \caption{Test set ROC curve with two events (average curve $\pm$ standard deviation intervals across five experimental repeats). The x-axis is on a log scale to emphasize the low FPR~regime.}
  \label{fig:roc-app-cr2}\vspace{-1.5em}
\end{figure}

We provide the ROC when we consider only two competing events, awakening and death not by withdrawal from life-sustaining therapies, at $t=6,12$ and $\Delta=48,72$, as shown in Figure~\ref{fig:roc-app-cr2}.

\end{document}